\documentclass[%
reprint,
superscriptaddress,
showpacs,showkeys,
 amsmath,amssymb,
 aps,
 pra,
]
{revtex4-1}

\usepackage{graphicx}
\usepackage{dcolumn}
\usepackage{bm}
\usepackage{mathrsfs}
\usepackage{extarrows}
\usepackage[colorlinks,citecolor=blue,linkcolor=blue]{hyperref}

\begin{document}

\preprint{APS/123-QED}

\title{Strong mechanical squeezing in a standard optomechanical system by pump modulation}

\author{Cheng-Hua Bai}
\affiliation{Department of Physics, Harbin Institute of Technology, Harbin, Heilongjiang 150001, China}%
\author{Dong-Yang Wang}%
\affiliation{Department of Physics, Harbin Institute of Technology, Harbin, Heilongjiang 150001, China}%
\author{Shou Zhang}
\email{szhang@ybu.edu.cn}
\affiliation{Department of Physics, Harbin Institute of Technology, Harbin, Heilongjiang 150001, China}
\affiliation{Department of Physics, College of Science, Yanbian University, Yanji, Jilin 133002, China}
\author{Shutian Liu}
\affiliation{Department of Physics, Harbin Institute of Technology, Harbin, Heilongjiang 150001, China}
\author{Hong-Fu Wang}
\email{hfwang@ybu.edu.cn}
\affiliation{Department of Physics, College of Science, Yanbian University, Yanji, Jilin 133002, China}




\date{\today}

\begin{abstract}
Being beneficial for the amplitude modulation of the pump laser, we propose a simple yet surprisingly effective mechanical squeezing scheme in a standard optomechanical system. By merely introducing a specific kind of periodic modulation into the single-tone driving field to cool down the mechanical Bogoliubov mode, the far beyond 3-dB strong mechanical squeezing can be engineered without requiring any additional techniques. Specifically, we find that the amount of squeezing is not simply dependent on the order of magnitude of the effective optomechanical coupling but strongly on the ratio of sideband strengths for it. To maximize the mechanical squeezing, we numerically and analytically optimize this ratio in the steady-state regime, respectively. The mechanical squeezing engineered in our scheme also has strong robustness and can survive at a high bath temperature. Compared with previous schemes based on the two-tone pump technique, our scheme involves fewer external control laser source and can be extended to other quantum systems to achieve strong squeezing effect.

\end{abstract}

\maketitle


\section{Introduction}\label{Sec1}
In recent years, with the enormous advances in optomechanics, including the experimental realization of quantum ground-state cooling for mechanical oscillators~\cite{2010Nature464697,2011Nature47889,2011Nature475359,2016PRL116063601,2017Nature541191,2019Nature56865} and exploitation of strong optomechanical coupling~\cite{2009Nature460724,2011Nature471204,2019arXiv191101262}, optomechanical system, the study of the controllable radiation-pressure interaction between the optical (microwave) and mechanical degrees of freedom, has been the flexible platform for the quantum manipulation of macroscopic mechanical oscillators in the fields of fundamental research and applied science~\cite{2014RMP861391,2008Science3211172,2012PhysToday6529,2013ADP525215}. Particularly, the exploration of the quantum-to-classical transition~\cite{2008Science3211172,1991PhysToday4436}, the search for novel quantum effects at macroscale~\cite{2005PhysToday5836}, and the pursuit to measure extremely weak signals (gravitational waves) with ultrahigh precision at the quantum level of sensitivity~\cite{1980RMP52341,1992Science256325} have been the primary thrust for the engineering strong mechanical squeezing in optomechanical systems. Therefore, many significant efforts have been devoted to developing alternative mechanical squeezing methods and techniques. In the early parametric amplification mechanical squeezing scheme, being similar to the parametric technique applied to optical squeezing~\cite{1981OC39401}, due to the limitation of the system instability, the amount of mechanical squeezing cannot be reduced below one-half of the standard quantum limit (i.e., the so-called 3-dB limit)~\cite{1991PRL67699}. Based on the cavity optomechanical system, many schemes are also proposed to generate mechanical squeezing, such as modulation of the external driving field~\cite{2009PRL103213603,2011PRA83033820,2019ADP1800271}, quantum squeezing transfer from the optical parametric amplifier to mechanical oscillator~\cite{2016PRA93043844}, and $XX$ type interaction induced by mechanical non-Markovian reservoir~\cite{2018OL436053}. Although above schemes possess respective advantages under certain circumstances, the achieved mechanical squeezing is relatively weak and fails to break the 3-dB limit. As a consequence, to overcome this limit, other strong mechanical squeezing schemes are proposed accordingly, including squeezed light driving and squeezing transfer~\cite{2009PRA79063819}, quadratic optomechanical coupling~\cite{2014PRA89023849}, dissipative optomechanical coupling~\cite{2010PRA82033811,2013PRA88013835}, and Duffing nonlinearity~\cite{2015PRA91013834,2017PRA96013837,2019PRA100033812}. We also investigated the joint mechanical squeezing effect between two different kinds of squeezing techniques instead of only one squeezing manipulation method in the above schemes and found that the beyond 3-dB limit strong mechanical squeezing can be easily engineered, but each kind of independent squeezing component is permitted below 3 dB~\cite{2019PRs71229}. Furthermore, some schemes even resorted to more complex techniques, such as quantum measurement~\cite{2008NJP10095010,2011PRL107213603,2013PRL110184301}, quantum feedback~\cite{2005PRB71235407}, modulations of radiation-pressure coupling and mechanical spring constant~\cite{2018OE26011915}, combination of both linear and quadratical optomechanical couplings and squeezed light injection~\cite{2018PRA97043619}, and simultaneous linear and nonlinear couplings and amplitude-modulated driving field~\cite{2017PRA96063811}.

In fact, another powerful approach to effectively manipulate quantum states is reservoir engineering. Due to the advantages of the independence of initial state for the system and the robustness with respect to decoherence for the environment, this technique is greatly high performance in experimental implementations and has been widely applied in cavity (circuit) quantum electrodynamics~\cite{2011PRA84064302,2012PRL109183602,2014PRA89013820,2014PRA90054302,2018PRA98042310} and cavity optomechanics~\cite{2013PRA87033829,2013PRL110253601,2019Nature570480,2014PRA89063805,2018Nature556478,2013PRA88063833,2015Science349952,2015PRL115243601,2016PRL117100801}. In Ref.~\cite{2013PRA87033829}, a stationary two-mode squeezed vacuum state of two mechanical oscillators can be generated by cavity dissipation. The highly entangled cavity fields can be achieved by mechanical dissipation in a three-mode optomechanical system where two optical or microwave cavity modes are coupled to a common mechanical mode~\cite{2013PRL110253601}. Very recently, the theoretical work in Ref.~\cite{2013PRL110253601} has been successfully demonstrated in experiment and the stationary emission of entangled microwave radiation fields can be observed~\cite{2019Nature570480}. However, Ref.~\cite{2014PRA89063805} considered a different case where two mechanical oscillators are independently coupled to a common cavity mode and the strong mechanical-mechanical entanglement can be prepared by engineering of a single reservoir. Based on this scheme, the stabilized entanglement between two massive micromechanical oscillators has also been reported experimentally using the technique of reservoir engineering~\cite{2018Nature556478}. The typical mechanical squeezing scheme applying the reservoir engineering technique into optomechanics is driving an optical or microwave cavity with a pair of pump tones at $\omega_c\mp\omega_m$ ($\omega_c$ is optical or microwave frequency while $\omega_m$ is mechanical frequency)~\cite{2013PRA88063833}, and there is a requirement that the red-detuned pump should be at a higher power than the blue-detuned pump. Subsequently, utilizing the reservoir engineering technique based on two-tone driving, some experimental works have manipulated micromechanical oscillator into a quantum squeezed state~\cite{2015Science349952,2015PRL115243601,2016PRL117100801}. Hence, a novel and interesting idea arises: Whether above schemes~\cite{2013PRA88063833,2015Science349952,2015PRL115243601,2016PRL117100801} can be well workable when there is only a single-tone pump? This curious question is we wish to address.

In this paper, we consider a standard optomechanical device involving only one cavity mode and one mechanical mode which are coupled through radiation-pressure interaction. A specific kind of periodic amplitude modulation is introduced into the single-tone driving field. This operation leads to a desired form of the effective optomechanical coupling in the long-time limit, which just permits to cool down the Bogoliubov mode of the mechanical mode via the interaction with the cavity mode. Under this mechanism, the far surpassing 3-dB strong mechanical squeezing can be engineered. We discuss in detail the effects of the nonresonant terms induced by the periodic structure of the effective optomechanical coupling on the mechanical squeezing, which results in the direction of quadrature squeezing rotates continuously in phase space. To maximize the squeezing, we numerically and analytically optimize the ratio for the effective optomechanical coupling sideband strengths, respectively, which balances the competing effect between two opposing tendencies at the largest degree. We also note that the engineered mechanical squeezing is robust against the mechanical thermal noise and can survive at a high bath temperature. Besides involving fewer external control laser source compared with the previous schemes, our scheme can also be generalized to simplify some existing schemes, such as dissipative generation of squeezed output field~\cite{2014NJP16063058} and mechanical squeezing in an unresolved-sideband regime~\cite{2019PRA99043805} based on the two pump tones.

The rest of this paper is organized as follows. In Sec.~\ref{Sec2}, we introduce the standard optomechanical system driven by a periodically amplitude-modulated single-tone pump field. In Sec.~\ref{Sec3}, we illustrate the periodic dynamics of the manipulated optomechanical system and derive the linear quantum Langevin equation. In Sec.~\ref{Sec4}, we obtain the required effective optomechanical coupling for the generation of mechanical squeezing and discuss in detail from the perspectives of nonresonant terms influence without rotating-wave approximation, balancing the competing effects, and optimizing ratio for the effective optomechanical coupling sideband strengths, respectively. In Sec.~\ref{Sec5}, via eliminating the cavity mode adiabatically, the explicitly analytical expressions about the stationary mechanical squeezing and the optimized ratio for the sideband strengths are acquired and the experimental feasibility is also briefly analyzed. Finally, we summarize our work in Sec.~\ref{Sec6}.

\section{Model and Hamiltonian}\label{Sec2}
\begin{figure}
\centering
\includegraphics[width=0.7\linewidth]{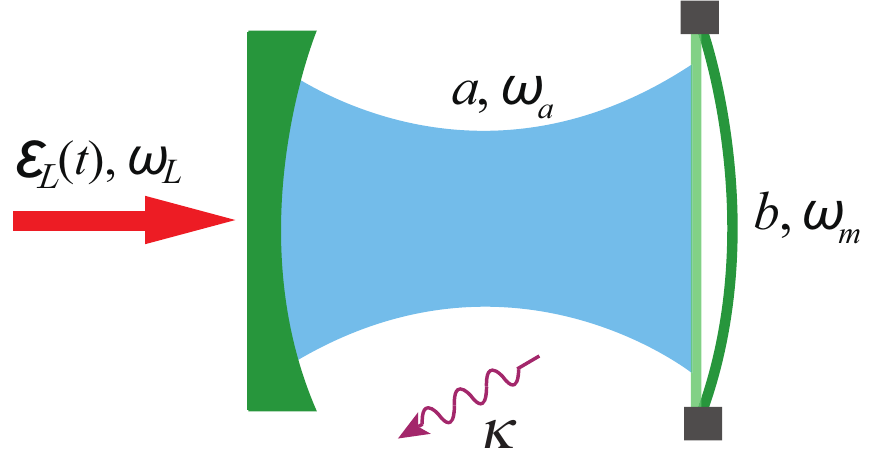}
\caption{(Color online) Schematic diagram of the optomechanical setup for achieving the arbitrarily strong mechanical squeezing, where the cavity field driven by a periodically amplitude-modulated external laser field couples to the mechanical mode via the controllable radiation-pressure interaction.}\label{Fig1}
\end{figure}

The sketch of the modulated optomechanical system is shown in Fig.~\ref{Fig1}, in which a periodically amplitude-modulated external driving field [with amplitude $\varepsilon_L(t)$ and frequency $\omega_L$] is imposed on the standard optomechanical system. Under the strong driving regime, the optical field (with frequency $\omega_a$ and decay rate $\kappa$)  interacts with the mechanical oscillator (with frequency $\omega_m$ and damping $\gamma_m$) via the manipulable radiation-pressure effect. In the rotating frame with respect to the laser frequency $\omega_L$, the system Hamiltonian can be written as $(\hbar=1)$
\begin{eqnarray}\label{Eq1}
H&=&\delta_aa^{\dag}a+\omega_mb^{\dag}b-g_0a^{\dag}a(b+b^{\dag}) \cr\cr
&&+i\left[\varepsilon_L(t)e^{i\varphi}a^{\dag}-\varepsilon_L^{\ast}(t)e^{-i\varphi}a\right],
\end{eqnarray}
where $\delta_a=\omega_a-\omega_L$ is the frequency detuning of the cavity with respect to the input laser, $a^{\dag}$ ($b^{\dag}$) and $a$ ($b$) are the creation and annihilation operators of cavity (mechanical) mode, respectively, and $g_0$ is the single-photon optomechanical coupling strength. $\varepsilon_L(t)$ is the periodically modulated amplitude of the external driving field and is carried out a modulation period $\tau$, i.e., $\varepsilon_L(t)=\varepsilon_L(t+\tau)=\sum_{n=-\infty}^{\infty}\varepsilon_ne^{-in\Omega t}$, in which $\Omega=2\pi/\tau$ is the modulation frequency and $\varepsilon_n$ is the sideband-modulation strength associated with the corresponding sideband power $P_n$ by $\varepsilon_n=\sqrt{2\kappa P_n/(\hbar\omega_L)}$. $\varphi$ is the phase of the laser field coupling to cavity mode $a$~\cite{2015PRA91053854}. For simplicity, we usually set $\varphi=0$ in generic optomechanical system~\cite{2007PRL98030405,2011PRL107063601}. Later we will show the extremely vital role of the modulation sidebands ($\sim e^{\pm i\Omega t}$) playing in engineering the arbitrarily strong mechanical squeezing.

Due to the coupling between the optomechanical system and the environment, the system dynamics will be inevitably influenced by the cavity decay and mechanical damping. Taking these dissipative elements into account, the quantum Langevin equations (QLEs) that dominate the system dynamical evolution are as follows
\begin{eqnarray}\label{Eq2}
\frac{da}{dt}&=&-i\delta_aa+ig_0a(b+b^{\dag})+\varepsilon_L(t)-\frac{\kappa}{2}a+\sqrt{\kappa}a_{\mathrm{in}}(t), \cr\cr
\frac{db}{dt}&=&-i\omega_mb+ig_0a^{\dag}a-\frac{\gamma_m}{2}b+\sqrt{\gamma_m}b_{\mathrm{in}}(t),
\end{eqnarray}
where $a_{\mathrm{in}}(t)$ and $b_{\mathrm{in}}(t)$ are, respectively, the zero-mean cavity vacuum input noise operator and mechanical thermal noise operator. Under the Markovian reservoir assumption, the nonzero correlation functions of the noise operators $a_{\mathrm{in}}$ and $b_{\mathrm{in}}$ are~\cite{Book1}
\begin{eqnarray}\label{Eq3}
\langle a_{\mathrm{in}}^{\dag}(t)a_{\mathrm{in}}(t^{\prime})\rangle&=&n_a\delta(t-t^{\prime}), \cr\cr
\langle a_{\mathrm{in}}(t)a_{\mathrm{in}}^{\dag}(t^{\prime})\rangle&=&(n_a+1)\delta(t-t^{\prime}), \cr\cr
\langle b_{\mathrm{in}}^{\dag}(t)b_{\mathrm{in}}(t^{\prime})\rangle&=&n_m\delta(t-t^{\prime}), \cr\cr
\langle b_{\mathrm{in}}(t)b_{\mathrm{in}}^{\dag}(t^{\prime})\rangle&=&(n_m+1)\delta(t-t^{\prime}),
\end{eqnarray}
where $n_a$ and $n_m$ are mean thermal occupancies of the cavity bath and mechanical bath, respectively.

\section{System periodic dynamics}\label{Sec3}
Strong external driving induces large amplitudes of the cavity mode and mechanical mode so that the standard linearization technology can be applied to the nonlinear QLEs in Eq.~(\ref{Eq2}). For this purpose, we write cavity mode $a$ and mechanical mode $b$ as the sum of the classical mean value and the quantum fluctuation operator, i.e., $\mathscr{O}\rightarrow\langle\mathscr{O}(t)\rangle+\mathscr{O} (\mathscr{O}=a, b)$. In this case, we obtain the set of equation of motion about $\langle a(t)\rangle$ and $\langle b(t)\rangle$
\begin{eqnarray}\label{Eq4}
\frac{d\langle a(t)\rangle}{dt}&=&-i\delta_a\langle a(t)\rangle+ig_0\langle a(t)\rangle\left[\langle b(t)\rangle+\langle b(t)\rangle^{\ast}\right] \cr\cr
&&+\varepsilon_L(t)-\frac{\kappa}{2}\langle a(t)\rangle, \cr\cr
\frac{d\langle b(t)\rangle}{dt}&=&-i\omega_m\langle b(t)\rangle+ig_0|\langle a(t)\rangle|^2-\frac{\gamma_m}{2}\langle b(t)\rangle.
\end{eqnarray}
The linearized QLEs for the quantum fluctuation operators can be correspondingly acquired
\begin{eqnarray}\label{Eq5}
\frac{da}{dt}&=&-i\Delta_aa+ig_0\langle a(t)\rangle(b+b^{\dag})-\frac{\kappa}{2}a+\sqrt{\kappa}a_{\mathrm{in}}(t), \cr\cr
\frac{db}{dt}&=&-i\omega_mb+ig_0\langle a(t)\rangle^{\ast}a+ig_0\langle a(t)\rangle a^{\dag}-\frac{\gamma_m}{2}b \cr\cr
&&+\sqrt{\gamma_m}b_{\mathrm{in}}(t), 
\end{eqnarray}
where $\Delta_a=\delta_a-g_0[\langle b(t)\rangle+\langle b(t)\rangle^{\ast}]$ is the effective cavity detuning sightly modulated by the mechanical motion.

Owing to the periodic modulation of the external driving acting on the optomechanical cavity [$\varepsilon_L(t)=\varepsilon_L(t+\tau)$], according to the Floquet theorem~\cite{Book2}, for the present linearized dynamical system, the cavity mode amplitude $\langle a(t)\rangle$ and mechanical mode amplitude $\langle b(t)\rangle$ will acquire the same modulation period with the performed external driving in the asymptotic regime, i.e., $\lim\limits_{t\rightarrow\infty}\langle a(t)\rangle=\langle a(t+\tau)\rangle$ and $\lim\limits_{t\rightarrow\infty}\langle b(t)\rangle=\langle b(t+\tau)\rangle$. 

For simplicity and to produce the desired system dynamics for generating mechanical squeezing, we need only to truncate the driving-modulation sidebands to the order of $e^{\pm i\Omega t}$, i.e., $\varepsilon_L(t)=\sum_{n=-1}^{^1}\varepsilon_ne^{-in\Omega t}$. As a result, the cavity mode amplitude $\langle a(t)\rangle$ and mechanical mode amplitude $\langle b(t)\rangle$ will have the same form with the chosen external driving-modulation structure in the long-time limit
\begin{eqnarray}\label{Eq6}
\langle\mathscr{O}(t)\rangle=\mathscr{O}_{-1}e^{i\Omega t}+\mathscr{O}_0+\mathscr{O}_1e^{-i\Omega t}~~(\mathscr{O}=a, b),
\end{eqnarray}
where $\mathscr{O}_n$ are the sideband amplitudes for the cavity and mechanical modes with $n=-1, 0, 1$. More details see Appendix~\ref{App1}.

\section{Engineering of mechanical squeezing}\label{Sec4}
In order to reveal the significantly important effect of the modulation sidebands ($\sim e^{\pm i\Omega t}$) on the engineering mechanical squeezing, we define $g_0\langle a(t)\rangle$ in Eq.~(\ref{Eq4}) as the effective optomechanical coupling $G(t)$ and specify it to 
\begin{eqnarray}\label{Eq7}
G(t)=g_0\langle a(t)\rangle=G_{-1}e^{i\Omega t}+G_0+G_1e^{-i\Omega t},
\end{eqnarray}
where $G_n$ ($n=-1, 0, 1$) are time-independent positive reals and associated with the driving sideband components $\varepsilon_n$. By further introducing the slow varying fluctuation operators with tildes $a=\tilde{a}e^{-i\Delta_at}$, $b=\tilde{b}e^{-i\omega_mt}$, $a_{\mathrm{in}}=\tilde{a}_{\mathrm{in}}e^{-i\Delta_at}$, $b_{\mathrm{in}}=\tilde{b}_{\mathrm{in}}e^{-i\omega_mt}$, setting the effective cavity detuning as the anti-Stokes sideband $\Delta_a=\omega_m$ and the external driving modulation frequency $\Omega=2\omega_m$, and assuming the effective optomechanical coupling sideband amplitudes are weak, i.e., $G_n\ll\omega_m$, the linearized QLEs for the operators $\tilde{a}$ and $\tilde{b}$ can be simplified as
\begin{eqnarray}\label{Eq8}
\dot{\tilde{a}}&=&iG_0\tilde{b}+iG_1\tilde{b}^{\dag}-\frac{\kappa}{2}\tilde{a}+\sqrt{\kappa}\tilde{a}_{\mathrm{in}}(t), \cr\cr
\dot{\tilde{b}}&=&iG_0\tilde{a}+iG_1\tilde{a}^{\dag}-\frac{\gamma_m}{2}\tilde{b}+\sqrt{\gamma_m}\tilde{b}_{\mathrm{in}}(t),
\end{eqnarray}
where the fast oscillating terms $e^{\pm2i\omega_mt}$ and $e^{\pm4i\omega_mt}$ have been omitted safely under the rotating-wave approximation (RWA) and whose nonresonant effects will be discussed later.

For convenience, we introduce the quadrature fluctuation operators with tildes
\begin{eqnarray}\label{Eq9}
\delta\widetilde{X}_{\mathscr{O}=a, b}&=&(\tilde{\mathscr{O}}+\tilde{\mathscr{O}}^{\dag})/\sqrt{2}, \cr\cr
\delta\widetilde{Y}_{\mathscr{O}=a, b}&=&(\tilde{\mathscr{O}}-\tilde{\mathscr{O}}^{\dag})/\sqrt{2}i,
\end{eqnarray}
and the quadrature noise operators with tildes
\begin{eqnarray}\label{Eq10}
\widetilde{X}_{\mathscr{O}=a,b}^{\mathrm{in}}&=&(\tilde{\mathscr{O}}_{\mathrm{in}}+\tilde{\mathscr{O}}_{\mathrm{in}}^{\dag})/\sqrt{2}, \cr\cr
\widetilde{Y}_{\mathscr{O}=a,b}^{\mathrm{in}}&=&(\tilde{\mathscr{O}}_{\mathrm{in}}-\tilde{\mathscr{O}}_{\mathrm{in}}^{\dag})/\sqrt{2}i.
\end{eqnarray}
Then, Eq.~(\ref{Eq8}) can be expressed in a more concise form:
\begin{eqnarray}\label{Eq11}
\dot{\widetilde{\mathrm{\textbf{R}}}}(t)=\widetilde{\mathrm{\textbf{M}}}\widetilde{\mathrm{\textbf{R}}}(t)+\widetilde{\mathrm{\textbf{N}}}(t),
\end{eqnarray}
where the vector $\widetilde{\mathrm{\textbf{R}}}$ about fluctuation operators is $\widetilde{\mathrm{\textbf{R}}}=[\delta\widetilde{X}_a, \delta\widetilde{Y}_a, \delta\widetilde{X}_b, \delta\widetilde{Y}_b]^T$, $4\times4$ time-independent coefficient matrix $\widetilde{\mathrm{\textbf{M}}}$ is
\begin{eqnarray}\label{Eq12}
\widetilde{\mathrm{\textbf{M}}}=
\begin{bmatrix}
-\frac{\kappa}{2} & 0 & 0 & -G_- \\\\
0 & -\frac{\kappa}{2} & G_+ & 0 \\\\
0 & -G_- & -\frac{\gamma_m}{2} & 0 \\\\
G_+ & 0 & 0 & -\frac{\gamma_m}{2}
\end{bmatrix},
\end{eqnarray}
and the noise operator victor $\widetilde{\mathrm{\textbf{N}}}$ is defined as $\widetilde{\mathrm{\textbf{N}}}=[\sqrt{\kappa}\widetilde{X}_a^{\mathrm{in}}, \sqrt{\kappa}\widetilde{Y}_a^{\mathrm{in}}, \sqrt{\gamma_m}\widetilde{X}_b^{\mathrm{in}}, \sqrt{\gamma_m}\widetilde{Y}_b^{\mathrm{in}}]^T$. Here $G_{\pm}=G_0\pm G_1$.

Obviously, Eq.~(\ref{Eq11}) which is completely equivalent to the linearized QLEs in Eq.~(\ref{Eq8}) is a first-order inhomogeneous differential equation with constant coefficient, whose formal solution can be written as
\begin{eqnarray}\label{Eq13}
\widetilde{\mathrm{\textbf{R}}}(t)=\widetilde{\mathrm{\textbf{G}}}(t)\widetilde{\mathrm{\textbf{R}}}(0)+
\widetilde{\mathrm{\textbf{G}}}(t)\int_0^t\widetilde{\mathrm{\textbf{G}}}^{-1}(\tau)\widetilde{\mathrm{\textbf{N}}}(\tau)d\tau,
\end{eqnarray}
in which $\widetilde{\mathrm{\textbf{G}}}(t)$ satisfies $\dot{\widetilde{\mathrm{\textbf{G}}}}(t)=\widetilde{\mathrm{\mathbf{M}}}\widetilde{\mathrm{\textbf{G}}}(t)$ and its initial condition is $\widetilde{\mathrm{\textbf{G}}}(0)=\mathrm{\mathbf{I}}$ (here $\mathrm{\mathbf{I}}$ is the identity matrix).

For more general regime of the optomechanical system, introducing the covariance matrix (CM) is more convenient for the study of the dynamical evolution of the system. To this end, we define a CM $\widetilde{\mathrm{\mathbf{V}}}(t)$ with components $\widetilde{\mathrm{\mathbf{V}}}_{ij}(t)=\langle\widetilde{\mathrm{\mathbf{R}}}_i(t)\widetilde{\mathrm{\mathbf{R}}}_j(t)\rangle$ for $i, j=1, 2, 3, 4$. Via further combining Eq.~(\ref{Eq13}), the explicit expression of the CM $\widetilde{\mathrm{\mathbf{V}}}(t)$ is
\begin{eqnarray}\label{Eq14}
\widetilde{\mathrm{\mathbf{V}}}(t)=\widetilde{\mathrm{\mathbf{G}}}(t)\widetilde{\mathrm{\mathbf{V}}}(0)\widetilde{\mathrm{\mathbf{G}}}^T(t)+
\widetilde{\mathrm{\mathbf{G}}}(t)\widetilde{\mathrm{\mathbf{S}}}(t)\widetilde{\mathrm{\mathbf{G}}}^T(t),
\end{eqnarray}
where
\begin{eqnarray}\label{Eq15}
\widetilde{\mathrm{\mathbf{S}}}(t)=\int_0^t\int_0^t\widetilde{\mathrm{\mathbf{G}}}^{-1}(\tau)\widetilde{\mathrm{\mathbf{K}}}(\tau, \tau^{\prime})
\big[\widetilde{\mathrm{\mathbf{G}}}^{-1}(\tau^{\prime})\big]^Td\tau d\tau^{\prime},
\end{eqnarray}
in which $\widetilde{\mathrm{\mathbf{K}}}(\tau, \tau^{\prime})$ is the so-called two-time noise correlation function whose elements are defined as $\widetilde{\mathrm{\mathbf{K}}}_{ij}(\tau, \tau^{\prime})=\langle\widetilde{\mathrm{\mathbf{N}}}_i(\tau)\widetilde{\mathrm{\mathbf{N}}}_j(\tau^{\prime})\rangle$. Obviously, the last two diagonal elements $\widetilde{\textbf{V}}_{33}(t)$ and $\widetilde{\textbf{V}}_{44}(t)$ of $\widetilde{\textbf{V}}(t)$ just are the variances for the mechanical position and momentum, respectively. Certainly, the degree of the mechanical squeezing can also be expressed in decibel units by $-10\log_{10}[\widetilde{\textbf{V}}_{jj}(t)/0.5]~(j=3, 4)$. Here, we specify that the cavity mode $a$ is prepared in vacuum state while the mechanical mode $b$ is in thermal state with the occupancy $n_m$ initially.

According to the Routh-Hurwitz stability criterion~\cite{1987PRA355288}, only if all eigenvalues of the time-independent coefficient matrix $\widetilde{\mathrm{\mathbf{M}}}$ in Eq.~(\ref{Eq11}) possess negative real parts, the system dynamics characterized by Eq.~(\ref{Eq11}) will be stable finally. For the current parameter regime, the stability constraint could be reduced as a simple form: $G_0>G_1$.

\subsection{Nonresonant effects without RWA}
In above discussion, we have ignored the nonresonant effects of the fast oscillating terms via making RWA. In this case, their functions in engineering mechanical squeezing are erased. To expose the contributions of the discarded high-frequency oscillating terms, we redefine the quadrature fluctuation operators, quadrature noise operators, and their corresponding operator vectors without tildes. Their forms are completely same with above definitions except the time-dependent coefficient matrix $\mathrm{\mathbf{M}}(t)$:
\begin{eqnarray}\label{Eq16}
\mathrm{\mathbf{M}}(t)=
\begin{bmatrix}
-\frac{\kappa}{2} & \Delta_a & -\mathrm{Im}[2G(t)] & 0 \\\\
-\Delta_a & -\frac{\kappa}{2} & \mathrm{Re}[2G(t)] & 0 \\\\
0 & 0 & -\frac{\gamma_m}{2} & \omega_m \\\\
\mathrm{Re}[2G(t)] & \mathrm{Im}[2G(t)] & -\omega_m & -\frac{\gamma_m}{2}
\end{bmatrix}, \cr
&&
\end{eqnarray}
where $\mathrm{Re}[\cdots]$ and $\mathrm{Im}[\cdots]$ indicate, respectively, the real and imaginary parts of a complex number. As a result, the fluctuation operator vector $\mathrm{\mathbf{R}}$ is
\begin{eqnarray}\label{Eq17}
\mathrm{\mathbf{R}}(t)=\mathrm{\mathbf{G}}(t)\mathrm{\mathbf{R}}(0)+\mathrm{\mathbf{G}}(t)\int_0^t\mathrm{\mathbf{G}}^{-1}(\tau)\mathrm{\mathbf{N}}(\tau)d\tau,
\end{eqnarray}
where $\mathrm{\mathbf{G}}(t)$ fulfills $\dot{\mathrm{\mathbf{G}}}(t)=\mathrm{\mathbf{M}}(t)\mathrm{\mathbf{G}}(t)$ and the initial condition is still $\mathrm{\mathbf{G}}(0)=\mathrm{\mathbf{I}}$.
\begin{figure}
\centering
\includegraphics[width=1.0\linewidth]{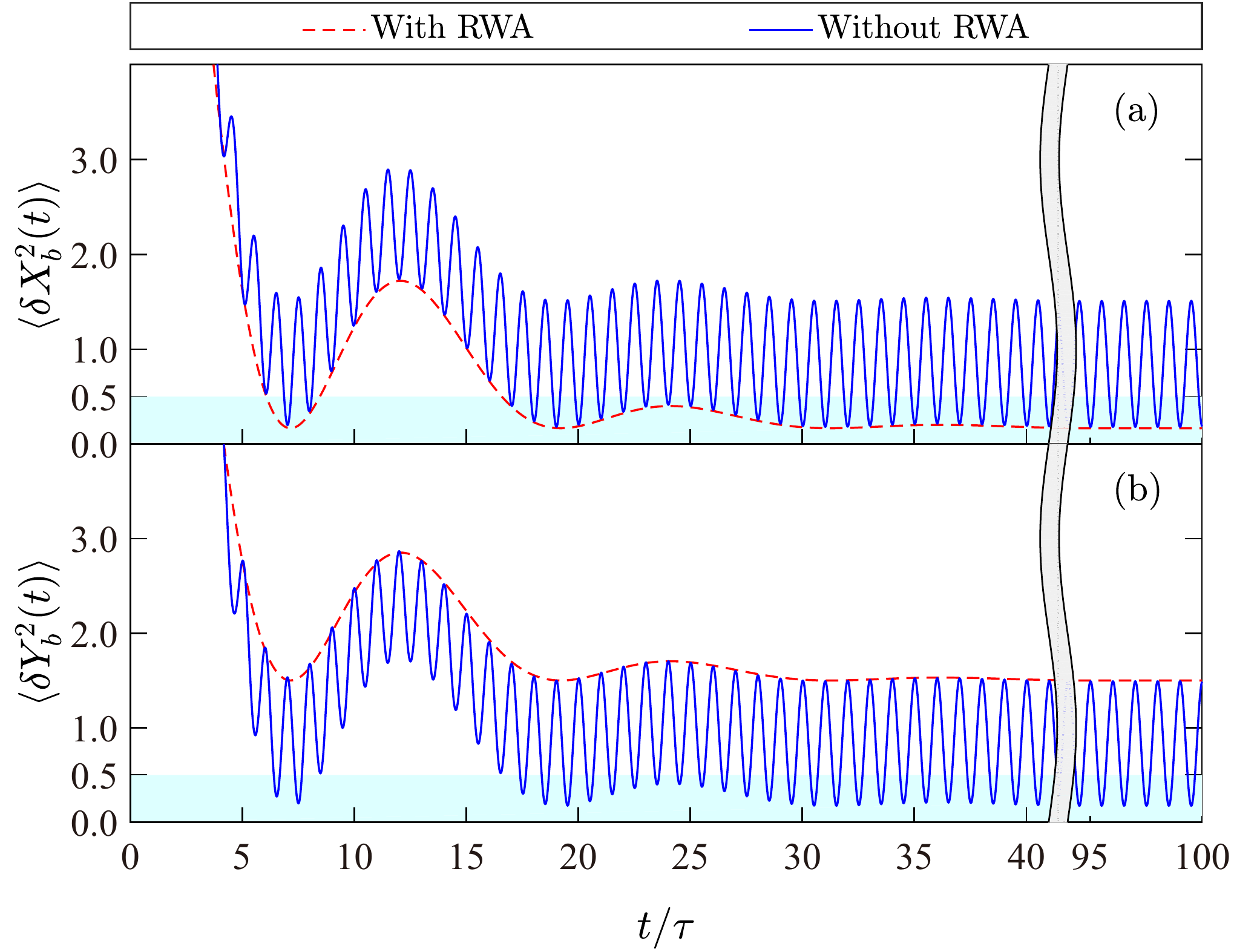}
\caption{(Color online) Time evolution of variances for the mechanical position and momentum fluctuations with and without RWA, respectively. The system parameters here we have set as: $\Delta_a=\omega_m$, $\kappa=0.1\omega_m$, $\gamma_m=10^{-6}\omega_m$, $G_{-1}=0.01\omega_m$, $G_0=0.1\omega_m$, $G_1=0.05\omega_m$, $n_m=10$, and $n_a=0$. The shadowed blue bottom region corresponds to mechanical squeezing.}\label{Fig2}
\end{figure}

To check the dynamics of quadrature squeezing explicitly, in Fig.~\ref{Fig2}, we display the time evolution of variances for the mechanical position and momentum fluctuations in the cases of with and without RWA in the modulation periods of $[0, 100\tau]$ when a typical parameter set of the effective optomechanical coupling sideband strengths $(G_{-1}, G_0, G_1)$ is given. Clearly, it is seen that, when the RWA is not made, the mechanical position and momentum will be periodically squeezed in the long-time limit and the squeezing period just is the performed external modulation period $\tau$. For example, in the modulation periods of $[95\tau, 100\tau]$, the mechanical position and momentum are all squeezed five times, respectively. However, as shown in Fig.~\ref{Fig2}, due to the bound of the Heisenberg uncertainty relation~\cite{Book3}, the mechanical position and momentum cannot be squeezed simultaneously. Once the RWA is exploited to erase the contributions of the high-frequency oscillating terms $e^{\pm2i\omega_mt}$ and $e^{\pm4i\omega_mt}$, the $\tau$-periodicity position and momentum squeezing will be collapsed to the direction of position and the momentum squeezing disappears accordingly. But no matter in the case of with or without RWA, the amount of mechanical squeezing is almost same during the entire evolution process.

On the other hand, the nonresonant effects of the fast oscillating terms can be revealed more intuitively in the phase space. To this end, it is necessary to introduce the Wigner function. Due to the above linearized system dynamics~[Eq.~(\ref{Eq5})] and the zero-mean Gaussian nature of the quantum noises, it ensures that the quantum steady state of the system is a Gaussian state~\cite{2007PRL98030405,2018PRA97022336}. Hence, as long as the CM is obtained, the Wigner function of the mechanical mode can be expressed as~\cite{Book4,2012RMP84621}.
\begin{eqnarray}\label{Eq18}
\mathcal{W}(\mathbf{D})=\frac{1}{2\pi\sqrt{\mathrm{Det}[\mathbf{V}_b]}}\mathrm{exp}\Big\{-\frac12\mathbf{D}^T\mathbf{V}_b^{-1}\mathbf{D}\Big\},
\end{eqnarray}
where $\mathbf{D}$ refers to the 2-dimensional vector $\mathbf{D}=[X_b, Y_b]^T$ and $\mathbf{V}_b$ is the CM for the mechanical mode.

\begin{figure*}
\centering
\includegraphics[width=1.0\linewidth]{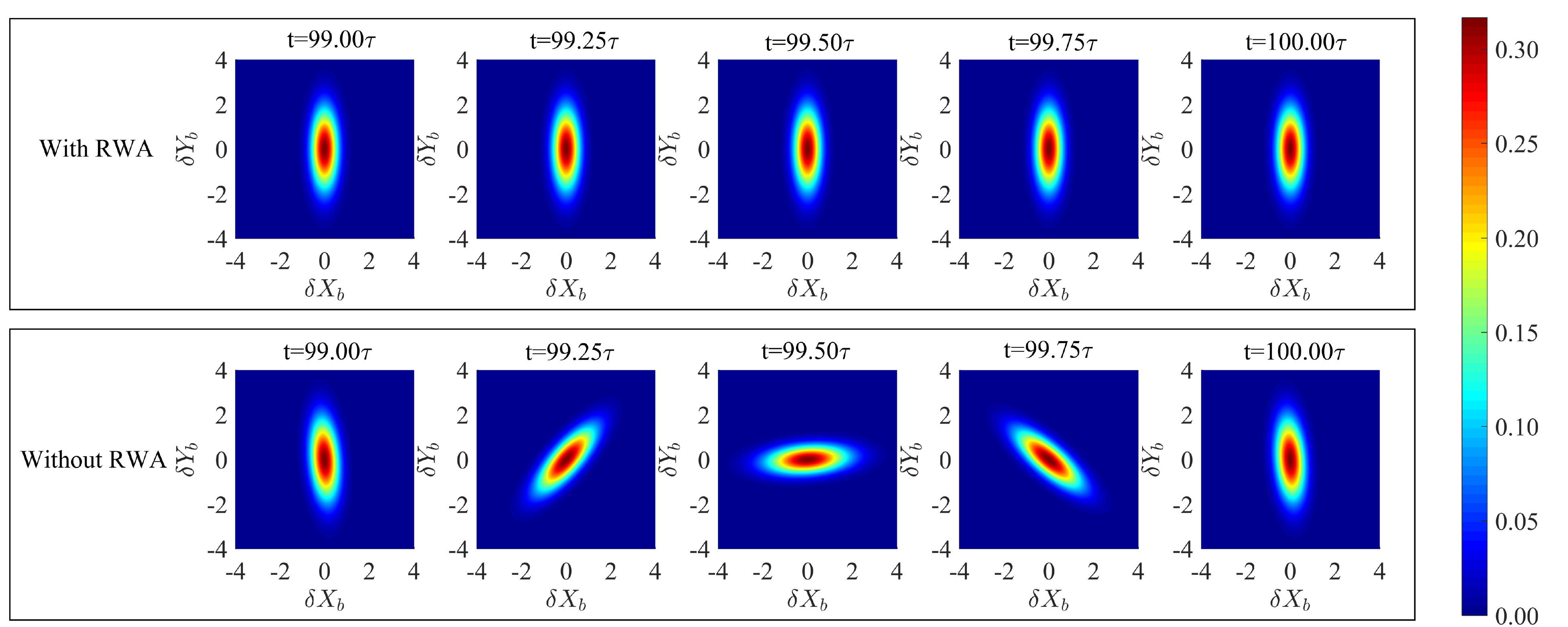}
\caption{(Color online) Wigner functions of mechanical mode at some different specific times in the cases of  with and without RWA. The system parameters are the same as in Fig.~\ref{Fig2}.}\label{Fig3}
\end{figure*}

In Fig.~\ref{Fig3}, we further show the Wigner functions of mechanical mode in the long-time modulation period of $[99\tau, 100\tau]$ for one quarter period time interval with and without RWA, respectively. One can observe from Fig.~\ref{Fig3} that, under the actions of the nonresonant effects of the fast oscillating terms, the direction of quadrature squeezing rotates continuously in phase space and the rotation period just corresponds to the modulation period $\tau$. This is becsuse the performed external driving is $\tau$ periodic [$\varepsilon_L(t)=\varepsilon_L(t+\tau)$] and according to the Floquet theory, the CM of the system will acquire the same periodicity of the external modulation in the long-time limit, i.e., $\widetilde{\mathrm{\textbf{V}}}(t)=\widetilde{\mathrm{\textbf{V}}}(t+\tau)$~\cite{2009PRL103213603,2012NJP14075014}. Therefore, from Eq.~(\ref{Eq18}), we can conclude that the Wigner function of the mechanical mode $\mathcal{W}$ will also satisfy $\mathcal{W}(\delta X_b, \delta Y_b, t)=\mathcal{W}(\delta X_b, \delta Y_b, t+\tau)$, which accounts for the period of rotation of the Wigner function in phase space is $2\tau$. However, when the high-frequency oscillating terms are omitted by RWA, the rotation of the Wigner functions at different specific times disappears and they all stretch along the vertical axis and contract along the horizontal axis, which clearly characterizes the mechanical squeezing in the direction of position. It is also found that, throughout all the Wigner functions no matter with or without RWA in Fig.~\ref{Fig3}, the shape of them is fixed, which again indicates that the degree of the mechanical squeezing is almost equivalent in two cases. This is since $G_0$ is maximum in parameter set ($G_{-1}, G_0, G_1$), the Stokes-scattering process $G_0e^{-2i\omega_mt}ab+G_0e^{2i\omega_mt}a^{\dag}b^{\dag}$ is the nearest resonant terms among the neglected high-frequency oscillating terms. In the low excitation of mechanical bath ($n_m=10$), the quantum backaction effect induced by the nearest resonant Stokes-scattering process on the mechanical mode is very weak. Therefore, the contribution of these high-frequency oscillating terms neglected by the RWA to the shape of Wigner function is not remarkable.

In addition, what we greatly concern is that, to achieve the desired form of the effective optomechanical coupling $G(t)$ as shown in Eq.~(\ref{Eq7}) for a given set of $(G_{-1}, G_0, G_1)$, we should how to choose a set of sideband-modulation strengths $(\varepsilon_{-1}, \varepsilon_0, \varepsilon_1)$ for the external driving $\varepsilon_L(t)$. See Appendix~\ref{App2} for more details.

In the present mechanical squeezing scheme, if we keep $G_0$ being fixed but add a $\pi$ phase to $G_1$, i.e., $G_1=|G_1|e^{i\pi}$, the quadrature squeezing about mechanical position and momentum fluctuations in the long-time modulation limit in Fig.~\ref{Fig2} will be reversed. As a result, the use of RWA leads also to the squeezing in the direction of momentum and the squeezing of position fluctuation vanishes accordingly.

\subsection{Competing effects between two opposing tendencies}
\begin{figure}
\includegraphics[width=1.0\linewidth]{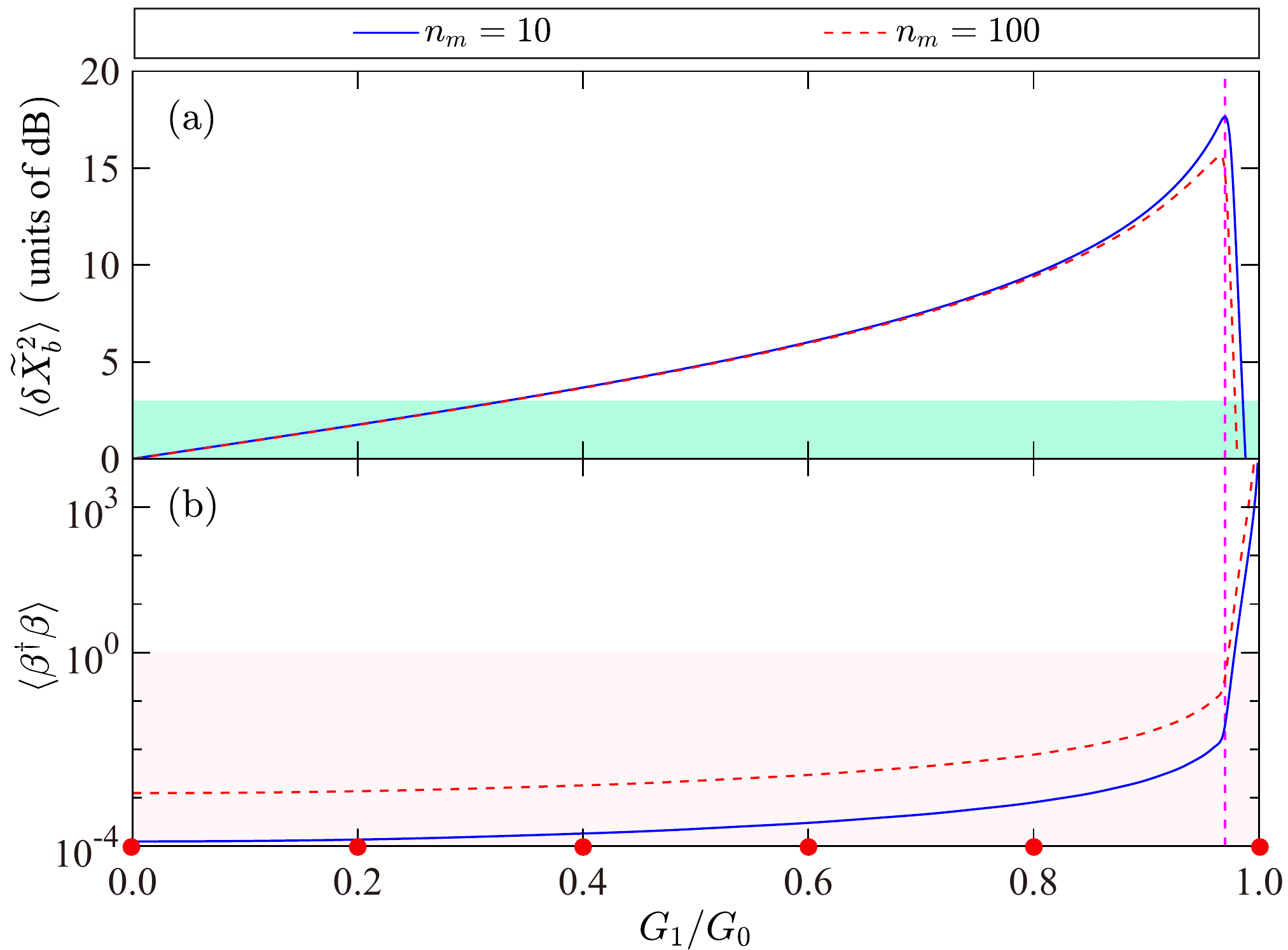}
\caption{(Color online) (a) Position variance $\langle\delta\widetilde{X}_b^2\rangle$ of the mechanical mode $b$ and (b) the occupancy $\langle\beta^{\dag}\beta\rangle$ for the Bogoliubov mode $\beta$ versus the ratio $G_1/G_0$ of the sideband strengths for the effective optomechanical coupling $G$ in the long-time limit. In both figures, the solid blue and dashed red curves correspond to the result obtained from $n_m=10$ and $n_m=100$, respectively. The shadowed green and pink bottom regions in (a) and (b) correspond, respectively, to mechanical squeezing below the 3 dB limit and Bogoliubov mode cooling for $\langle\beta^{\dag}\beta\rangle<1$. The dashed magenta vertical line corresponds to the position of the maximal $\langle\delta\widetilde{X}_b^2\rangle$ (dB) in the range of $G_1\in[0, G_0]$. The system parameters are the same as in Fig.~\ref{Fig2}.}\label{Fig4}
\end{figure}

In this subsection, we present the further interpretation of engineering mechanical squeezing via the dissipation of the cavity mode. For this reason, we introduce the Bogoliubov mode $\beta=\cosh r\tilde{b}+\sinh r\tilde{b}^{\dag}$ with $\tanh r=G_1/G_0$. In terms of the Bogoliubov mode, the QLEs in Eq.~(\ref{Eq8}) become
\begin{eqnarray}\label{Eq19}
\dot{\tilde{a}}&=&-\frac{\kappa}{2}\tilde{a}+i\mathcal{G}\beta+\sqrt{\kappa}\tilde{a}_{\mathrm{in}}(t), \cr\cr
\dot{\beta}&=&i\mathcal{G}\tilde{a}-\frac{\gamma_m}{2}\beta+\sqrt{\gamma_m}\beta_{\mathrm{in}}(t),
\end{eqnarray}
where $\mathcal{G}=\sqrt{G_0^2-G_1^2}$ is the effective coupling between the Bogoliubov mode and cavity mode and  $\beta_{\mathrm{in}}(t)=\cosh r\tilde{b}_{\mathrm{in}}(t)+\sinh r\tilde{b}_{\mathrm{in}}^{\dag}(t)$ is the effective noise corresponding to the Bogoliubov mode.

In Fig.~\ref{Fig4}, we plot the position variance $\langle\delta\widetilde{X}_b^2\rangle$ of the mechanical mode and the occupancy $\langle\beta^{\dag}\beta\rangle$ of the Bogoliubov mode as functions of the effective optomechanical coupling sideband strength ratio $G_1/G_0$ for a fixed $G_0$ in the case of different mean bath phonon numbers. It can be found that, in both cases of $n_m=10$ and $n_m=100$, with the increase of $G_1$ until a critical value, $\langle\delta\widetilde{X}_b^2\rangle$ is a monotonic function of $G_1/G_0$ and the mechanical squeezing becomes more and more stronger. In this corresponding range of $G_1/G_0$, the occupancy $\langle\beta^{\dag}\beta\rangle$ for the Bogoliubov mode is rising but very gently and the Bogoliubov mode remains in the ground-state cooling zone. However, with the continuous increase of $G_1$ once beyond this critical value, $\langle\delta\widetilde{X}_b^2\rangle$ declines rapidly but $\langle\beta^{\dag}\beta\rangle$ sharply grows on the contrary. One can also clearly see that, for this point of specific $G_1/G_0$, $\langle\delta\widetilde{X}_b^2\rangle$ takes the maximum and $\langle\beta^{\dag}\beta\rangle$ begins to sharply grow simultaneously. This kind of interesting competing effect can be throughly understood as follows.

According to Eq.~(\ref{Eq19}), in terms of the Bogoliubov mode $\beta$, the system Hamiltonian becomes
\begin{eqnarray}\label{Eq20}
\mathcal{H}=-\mathcal{G}(\tilde{a}\beta^{\dag}+\mathrm{H.c.}).
\end{eqnarray}
It shows that, obviously, the cavity mode $\tilde{a}$ and the Bogoliubov mode $\beta$ are coupled via the well-known beam-splitter Hamiltonian, which is usually applied to the optomechanical sideband cooling schemes of the mechanical mode~\cite{2015PRA91013824,2018PRA98023860,2018PRA98023816}. Therefore, the Bogoliubov mode $\beta$ can be cooled into ground state via the interaction with the cavity mode $\tilde{a}$. With the increase of $G_1$ for a fixed $G_0$, the squeezing parameter $r=\mathrm{arctanh}[G_1/G_0]$ will be enlarged accordingly. Hence,  as shown in Fig.~\ref{Fig4}(a), the mechanical mode is squeezed more strongly. On the other hand, with the continuous increase of $G_1$, the effective coupling between the cavity mode and the Bogoliubov mode $\mathcal{G}=\sqrt{G_0^2-G_1^2}$ will be decreased for a fixed $G_0$ and finally vanishes, which inhibits the ground-state cooling of the Bogoliubov mode more and more remarkably. So, as shown in Fig.~\ref{Fig4}(b), the occupancy $\langle\beta^{\dag}\beta\rangle$ rises gently at first and then grows sharply at last. Once the Bogoliubov mode $\beta$ cannot be cooled close to its ground state, the deleterious effect of the thermal noise will be in a dominant role and the amount of mechanical squeezing decreases quickly and disappears ultimately. Thus, the strongest mechanical squeezing for a fixed $G_0$ just is the balanced result from the competing effect of these two different kinds of opposing tendencies. The above novel phenomena verify again the fact that the cooling is a prerequisite to reveal the macroscopic quantum effects about the mechanical mode.

\subsection{Optimal ratio for the effective optomechanical coupling sideband strengths}
\begin{figure}
\centering
\includegraphics[width=1.0\linewidth]{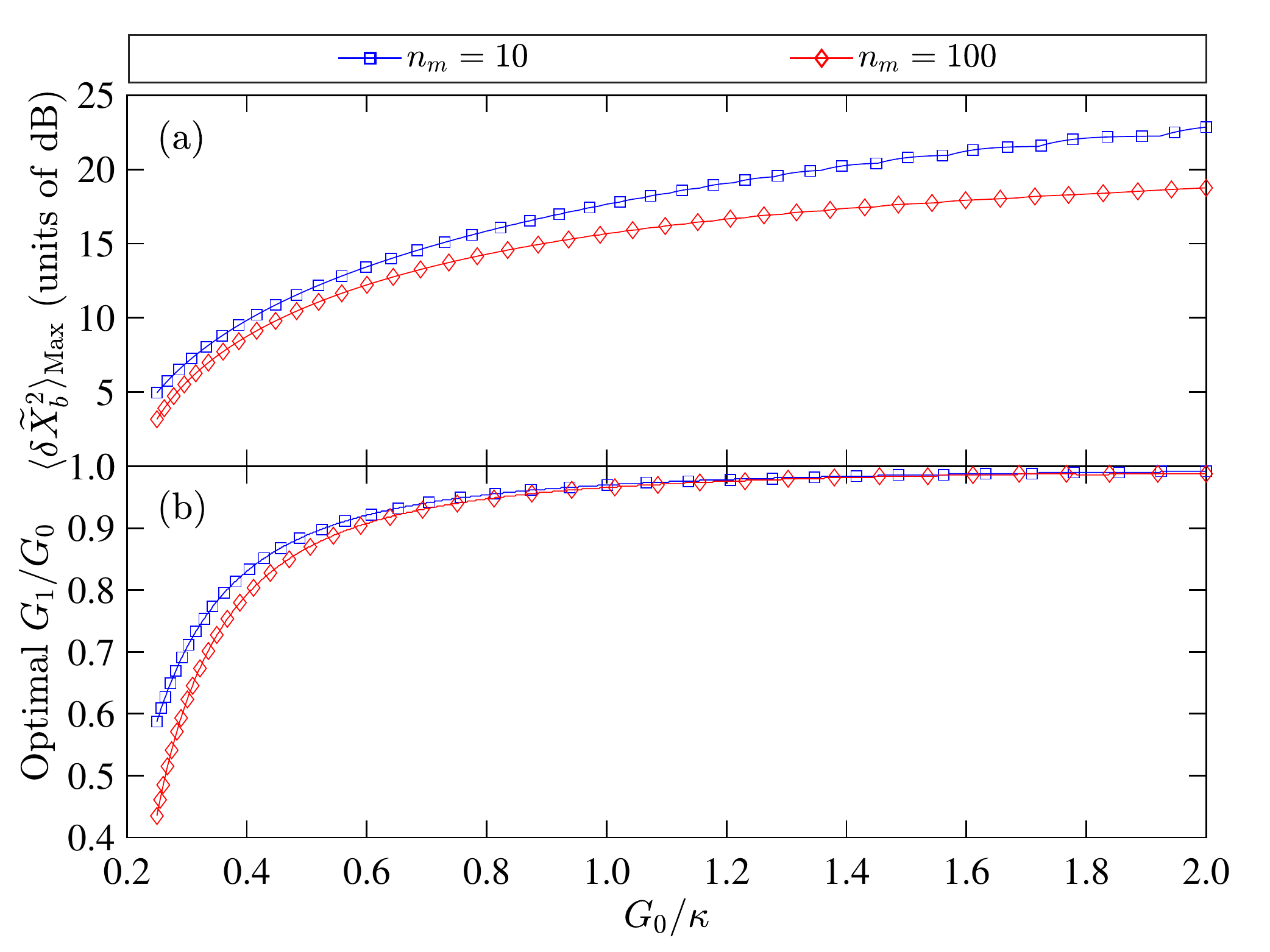}
\caption{(Color online) (a) The maximized position variance $\langle\delta\widetilde{X}_b^2\rangle$ of the mechanical mode and (b) the optimal ratio $G_1/G_0$ as functions of the effective optomechanical coupling center sideband strength $G_0$ for different mean thermal occupancies of the mechanical bath. Here the maximized $\langle\delta\widetilde{X}_b^2\rangle$ in (a) is acquired utilizing the corresponding optimal radio $G_1/G_0$ in (b) for different $G_0$. The system parameters are the same as in Fig.~\ref{Fig2}.}\label{Fig5}
\end{figure}

As illustrated in Fig.~\ref{Fig4}, for a fixed $G_0$, there is a specific $G_1$ that makes sure to maximize the mechanical squeezing. If $G_1$ is small, the mechanical squeezing is weak ($r$ is small). However, when $G_1$ is too large, the cooling capacity of the cavity mode is restrained significantly. Therefore, to engineer strong mechanical squeezing, it is very necessary to optimize the ratio $G_1/G_0$ over an appropriate range of $G_0$.

To this end, we numerically optimize the mechanical position variance $\langle\delta\widetilde{X}_b^2\rangle$ and the maximized $\langle\delta\widetilde{X}_b^2\rangle$ as functions of $G_0$ with different mechanical bath mean occupancies is shown in Fig.~\ref{Fig5}(a). Meanwhile, the corresponding optimal ratio $G_1/G_0$ which balances the competing effect between squeezing and cooling best for every $G_0$ is also presented in Fig.~\ref{Fig5}(b). As expected, from Fig.~\ref{Fig5}(a) one can note that, due to the adverse effect of the mechanical thermal noise, this is a reverse dependence of the maximized $\langle\delta\widetilde{X}_b^2\rangle$ on the mean thermal occupancy of the mechanical bath. On the other hand, with the increase of $G_0$, the coupling $\mathcal{G}=\sqrt{G_0^2-G_1^2}=G_0\sqrt{1-(G_1/G_0)^2}$ will be enhanced accordingly for a specific $G_1/G_0$. As a result, the cooling behavior performed by the cavity mode is more powerful, which induces that, as demonstrated in Fig.~\ref{Fig5}(b), the tendency of the optimal ratio $G_1/G_0$ is more and more close to unit but cannot equal to unit. It implies in turn the more stronger mechanical squeezing displayed in Fig.~\ref{Fig5}(a).

\begin{figure}
\centering
\includegraphics[width=0.7\linewidth]{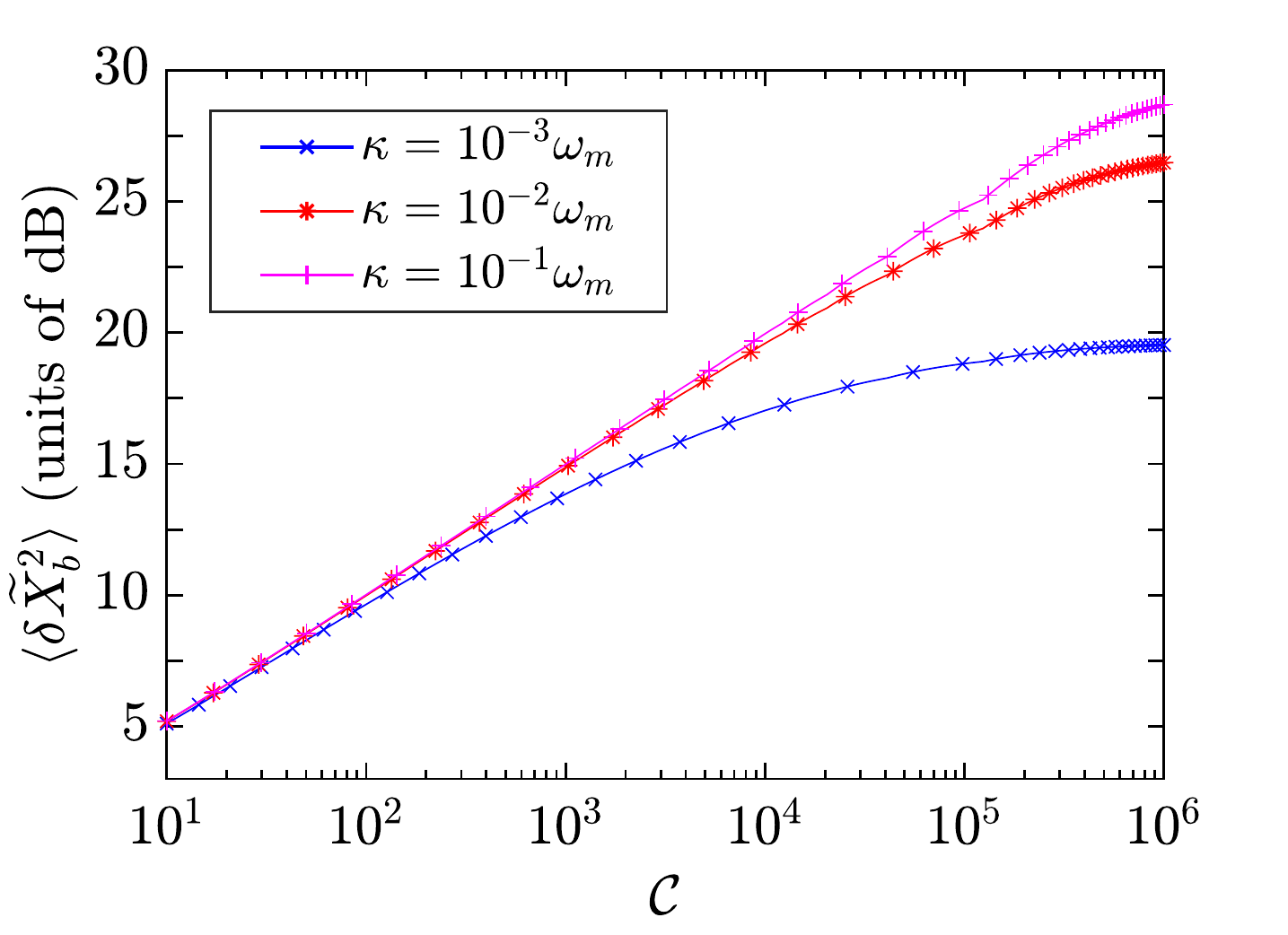}
\caption{(Color online) Position variance $\langle\delta\widetilde{X}_b^2\rangle$ of the mechanical mode versus the system cooperativity $\mathcal{C}$ with different cavity mode dissipation rate $\kappa$, in which $\langle\delta\widetilde{X}_b^2\rangle$ is obtained with the numerically optimized ratio $G_1/G_0$. Here the system parameters are chosen as $\gamma_m=10^{-5}\omega_m$ and $n_m=0$. Other parameters are the same as in Fig.~\ref{Fig2}.}\label{Fig6}
\end{figure}

In addition, it is well known that, the ground-state cooling of the Bogoliubov mode $\beta$ is not only dependent on the coupling strength $\mathcal{G}$ between the cavity mode and the Bogoliubov mode, but also related closely to the decay rate of the cavity mode itself. To shed light on the effect of the cavity mode decay rate in engineering mechanical squeezing clearly, in Fig.~\ref{Fig6}, we plot the position variance $\langle\delta\widetilde{X}_b^2\rangle$ as functions of the system cooperativity $\mathcal{C}=4G_0^2/(\kappa\gamma_m)$ with different decay rate $\kappa$, where $\langle\delta\widetilde{X}_b^2\rangle$ has been maximized by numerically optimizing the ratio over entire $G_1/G_0$. One can find that, the more stronger mechanical squeezing can be engineered in the limit of large system cooperativity $\mathcal{C}$. On the other hand, it is also clearly shown that the increasing decay rate of the cavity mode leads to a more stronger mechanical squeezing.

\section{Analytical solution for the steady-state mechanical squeezing}\label{Sec5}
In the present scheme, although we make use of the time-modulated input field to drive the optomechanical system [essentially, it is time modulated for the effective optomechanical coupling $G(t)$], as shown in Sec.~\ref{Sec4}, the time-dependent system dynamics [Eq.~(\ref{Eq5})] can be successfully transformed into the time-independent effective system dynamics [Eq.~(\ref{Eq19})] via making the RWA. Then based on the time-independent effective system dynamics after the RWA, as long as the condition of adiabatic approximation is satisfied, i.e., the cavity decay rate $\kappa$ is much larger than the effective coupling $\mathcal{G}$ between the cavity mode $\tilde{a}$ and the Bogoliubov mode $\beta$ ($\kappa\gg\mathcal{G}$), the cavity mode $\tilde{a}$ still can be adiabatically eliminated from the dynamics~\cite{2016PRA93043844}.

In this section, to better understand the mechanical squeezing effect and obtain the explicit optimal ratio of $G_1/G_0$, we analytically solve the position variance of the mechanical mode in the steady regime. From Eq.~(\ref{Eq19}), we obtain
\begin{eqnarray}\label{Eq26}
\tilde{a}\simeq\frac{2i\mathcal{G}}{\kappa}\beta+\frac{2}{\sqrt{\kappa}}\tilde{a}_{\mathrm{in}}(t),
\end{eqnarray}
and substitute which into Eq.~(\ref{Eq19}), we have
\begin{eqnarray}\label{Eq27}
\dot{\beta}\simeq-h\beta+\frac{2i\mathcal{G}}{\sqrt{\kappa}}\tilde{a}_{\mathrm{in}}(t)+\sqrt{\gamma_m}\beta_{\mathrm{in}}(t),
\end{eqnarray}
where $h=2\mathcal{G}^2/\kappa+\gamma_m/2$. From Eq.~(\ref{Eq27}), the dynamical equation for the position fluctuation operator $\delta Q_{\beta}$ of the Bogoliubov mode can be acquired
\begin{eqnarray}\label{Eq28}
\delta\dot{Q}_{\beta}=-h\delta Q_{\beta}+\mathcal{F}_1(t)+\mathcal{F}_2(t),
\end{eqnarray}
where
\begin{eqnarray}\label{Eq29}
\mathcal{F}_1(t)&=&-\frac{2\mathcal{G}}{\sqrt{\kappa}}\widetilde{Y}_a^{\mathrm{in}}(t), \cr\cr
\mathcal{F}_2(t)&=&\sqrt{\frac{\gamma_m}{2}}\left[\beta_{\mathrm{in}}(t)+\beta^{\dag}_{\mathrm{in}}(t)\right],
\end{eqnarray}
are the effective quantum Langevin forces acted on the Bogoliubov mode and their correlation functions are
\begin{eqnarray}\label{Eq30}
\langle\mathcal{F}_1(t)\mathcal{F}_1(t^{\prime})\rangle&=&\frac{4\mathcal{G}^2}{\kappa}(n_a+\frac12)\delta(t-t^{\prime}),\cr\cr
\langle\mathcal{F}_2(t)\mathcal{F}_2(t^{\prime})\rangle&=&\gamma_me^{2r}(n_m+\frac12)\delta(t-t^{\prime}).
\end{eqnarray}
According to Eqs.~(\ref{Eq28}) and (\ref{Eq30}), the dynamical equation about $\langle\delta Q_{\beta}^2\rangle$ is
\begin{eqnarray}\label{Eq31}
\frac{d}{dt}\langle\delta Q_{\beta}^2\rangle=-2h\langle\delta Q_{\beta}^2\rangle+\frac{4\mathcal{G}^2}{\kappa}(n_a+\frac12)+\gamma_me^{2r}(n_m+\frac12), \cr\cr
&&
\end{eqnarray}
and therefore the analytical solution of $\langle\delta Q_{\beta}^2\rangle$ in the steady-state regime is
\begin{eqnarray}\label{Eq32}
\langle\delta Q_{\beta}^2\rangle_{\mathrm{s}}=\frac{2\mathcal{G}^2}{h\kappa}(n_a+\frac12)+\frac{\gamma_m}{2h}e^{2r}(n_m+\frac12).
\end{eqnarray}
As a result, the analytical solution for the steady-state position variance $\langle\delta\widetilde{X}_b^2\rangle_{\mathrm{s}}$ of the mechanical mode can be obtained accordingly
\begin{eqnarray}\label{Eq33}
\langle\delta\widetilde{X}_b^2\rangle_{\mathrm{s}}&=&e^{-2r}\langle\delta Q_{\beta}^2\rangle_{\mathrm{s}} \cr\cr
&=&\frac{2\mathcal{G}^2}{h\kappa}e^{-2r}(n_a+\frac12)+\frac{\gamma_m}{2h}(n_m+\frac12).
\end{eqnarray}

Here, we consider two limit cases. When $G_1\rightarrow 0$, we obtain $\mathcal{G}\rightarrow G_0$, $r=\mathrm{arctanh}G_1/G_0\rightarrow 0$, and $h\rightarrow2G_0^2/\kappa+\gamma_m/2\simeq2G_0^2/\kappa$. Therefore,
\begin{eqnarray}\label{Eq34}
\lim_{G_1\rightarrow 0}\langle\delta\widetilde{X}_b^2\rangle_{\mathrm{s}}=(n_a+\frac12)+\frac{\kappa\gamma_m}{4G_0^2}(n_m+\frac12).
\end{eqnarray}
Under the conditions of the high-frequency optical bath ($n_a=0$) and the large system cooperativity $\mathcal{C}$, $\lim_{G_1\rightarrow 0}\langle\delta\widetilde{X}_b^2\rangle\simeq\frac12$ (0 dB), which indicates that the mechanical mode is in the vacuum state approximately and it greatly coincides with the case of $G_1\rightarrow 0$ in Fig.~\ref{Fig4}(a). Obviously, while for $G_1\rightarrow G_0$, we obtain $\mathcal{G}\rightarrow 0$, $r=\mathrm{arctanh}G_1/G_0\rightarrow\infty$, and $h\rightarrow\gamma_m/2$. Hence,
\begin{eqnarray}\label{Eq35}
\lim_{G_1\rightarrow G_0}\langle\delta\widetilde{X}_b^2\rangle_{\mathrm{s}}=n_m+\frac12,
\end{eqnarray}
which means that the cooling effect disappears completely and the mechanical mode is in a thermal state. It also matches very well for the situation of $G_1\rightarrow G_0$ in Fig.~\ref{Fig4}.

To check the accuracy of the analytical solution in Eq.~(\ref{Eq33}) obtained under the adiabatic approximation, we now turn to solve the exact numerical solution for the steady-state position variance $\langle\delta\widetilde{X}_b^2\rangle_{\mathrm{s}}$ of the mechanical mode. Taking the Fourier transform in both sides of Eq.~(\ref{Eq11}) by $f(t)=\frac{1}{2\pi}\int_{-\infty}^{\infty}f(\omega)e^{-i\omega t}d\omega$ and solving it in the frequency domain, we get the expression about the position fluctuation of the mechanical mode
\begin{eqnarray}\label{Eq36}
\delta\widetilde{X}_b(\omega)&=&A(\omega)\widetilde{X}_a^{\mathrm{in}}(\omega)+B(\omega)\widetilde{Y}_a^{\mathrm{in}}(\omega)+ \cr\cr
&&E(\omega)\widetilde{X}_b^{\mathrm{in}}(\omega)+F(\omega)\widetilde{Y}_b^{\mathrm{in}}(\omega),
\end{eqnarray}
where
\begin{eqnarray}\label{Eq37}
&&A(\omega)=0, B(\omega)=-\frac{4G_-\sqrt{\kappa}}{4G_-G_++(\gamma_m-2i\omega)(\kappa-2i\omega)}, \cr\cr
&&E(\omega)=\frac{2(\kappa-2i\omega)\sqrt{\gamma_m}}{4G_-G_++(\gamma_m-2i\omega)(\kappa-2i\omega)}, F(\omega)=0. \cr\cr
&&
\end{eqnarray}
Apparently, the contribution of the first two terms in Eq.~(\ref{Eq36}) originates from the optical bath vacuum input noise, while the last two terms correspond to the contribution of the mechanical bath thermal noise. When the effective optomechanical coupling sideband strengths satisfy $G_1=G_0$, $\delta\widetilde{X}_b(\omega)=\frac{\sqrt{\gamma_m}}{\frac{\gamma_m}{2}-i\omega}\widetilde{X}_b^{\mathrm{in}}(\omega)$. Not surprisingly, it shows that the mechanical oscillator will make quantum Brownian motion because of the coupling with the bath environment.

The correlation functions of the noise operators in Eq.~(\ref{Eq36}) are
\begin{eqnarray}\label{Eq38}
&&\langle\widetilde{X}_a^{\mathrm{in}}(\omega)\widetilde{X}_a^{\mathrm{in}}(\Omega)\rangle=
\langle\widetilde{Y}_a^{\mathrm{in}}(\omega)\widetilde{Y}_a^{\mathrm{in}}(\Omega)\rangle \cr\cr
&&~~~~~~~~~~~~~~~~~~~~~=(n_a+\frac12)2\pi\delta(\omega+\Omega), \cr\cr
&&\langle\widetilde{X}_a^{\mathrm{in}}(\omega)\widetilde{Y}_a^{\mathrm{in}}(\Omega)\rangle=
-\langle\widetilde{Y}_a^{\mathrm{in}}(\omega)\widetilde{X}_a^{\mathrm{in}}(\Omega)\rangle=
i\pi\delta(\omega+\Omega), \cr\cr
&&\langle\widetilde{X}_b^{\mathrm{in}}(\omega)\widetilde{X}_b^{\mathrm{in}}(\Omega)\rangle=
\langle\widetilde{Y}_b^{\mathrm{in}}(\omega)\widetilde{Y}_b^{\mathrm{in}}(\Omega)\rangle \cr\cr
&&~~~~~~~~~~~~~~~~~~~~~=(n_m+\frac12)2\pi\delta(\omega+\Omega), \cr\cr
&&\langle\widetilde{X}_b^{\mathrm{in}}(\omega)\widetilde{Y}_b^{\mathrm{in}}(\Omega)\rangle=
-\langle\widetilde{Y}_b^{\mathrm{in}}(\omega)\widetilde{X}_b^{\mathrm{in}}(\Omega)\rangle=
i\pi\delta(\omega+\Omega), \cr\cr
	&&
\end{eqnarray}
and the position fluctuation spectrum of the mechanical mode is defined as
\begin{eqnarray}\label{Eq39}
&&2\pi S_{\widetilde{X}_b}(\omega)\delta(\omega+\Omega) \cr\cr
&=&\frac12[\langle\delta\widetilde{X}_b(\omega)\delta\widetilde{X}_b(\Omega)\rangle+
\langle\delta\widetilde{X}_b(\Omega)\delta\widetilde{X}_b(\omega)\rangle].
\end{eqnarray}
Resorting to Eq.~(\ref{Eq38}), the position fluctuation spectrum $S_{\widetilde{X}_b}$ can be obtained
\begin{eqnarray}\label{Eq40}
S_{\widetilde{X}_b}(\omega)&=&\left[A(\omega)A(-\omega)+B(\omega)B(-\omega)\right](n_a+\frac12)+ \cr\cr
&&\left[E(\omega)E(-\omega)+F(\omega)F(-\omega)\right](n_m+\frac12).
\end{eqnarray}
In the case of $G_1=G_0$, the position fluctuation spectrum is simplified as $S_{\widetilde{X}_b}(\omega)=\gamma_m(n_m+\frac12)/(\frac{\gamma_m^2}{4}+\omega^2)$, which obviously represents a Lorentzian spectrum with single peak located at frequency zero and full width $\gamma_m$ at half maximum. The steady-state position variance $\langle\delta\widetilde{X}_b^2\rangle_{\mathrm{s}}$ can be calculated by
\begin{eqnarray}\label{eqnarray}\label{Eq41}
\langle\delta\widetilde{X}_b^2\rangle_{\mathrm{s}}=\frac{1}{2\pi}\int_{-\infty}^{\infty}S_{\widetilde{X}_b}(\omega)d\omega.
\end{eqnarray}
Under the condition of $G_1=G_0$, we find $\langle\delta\widetilde{X}_b^2\rangle_{\mathrm{s}}=n_m+\frac12$, which just is the case of analytical solution in Eq.~(\ref{Eq35}).

\begin{figure}
\centering
\includegraphics[width=1.0\linewidth]{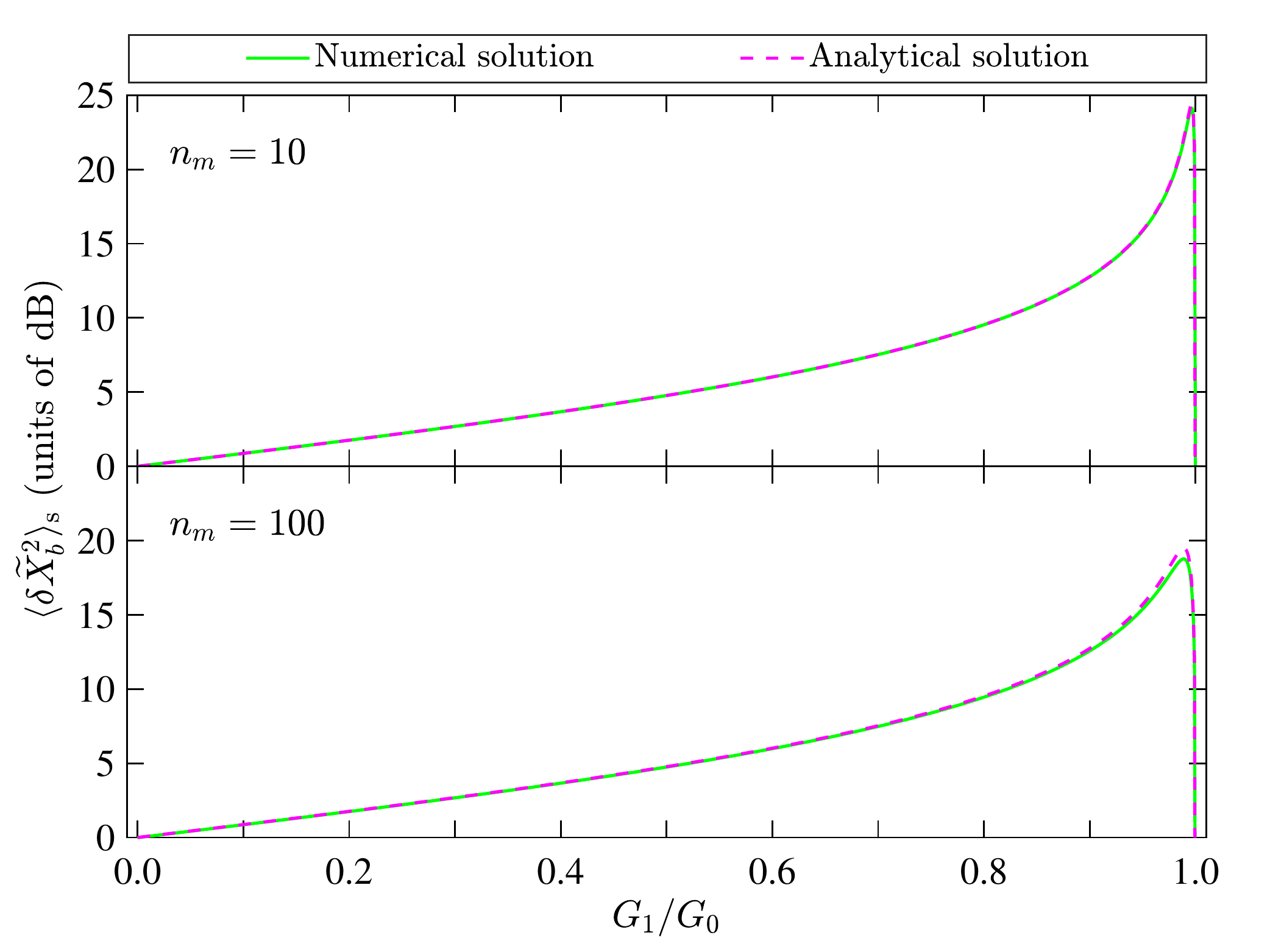}
\caption{(Color online) Comparison between the exact numerical solution and the approximate analytical solution for the steady-state position variance $\langle\delta\widetilde{X}_b^2\rangle_{\mathrm{s}}$ of the mechanical mode with different mechanical bath mean phonon numbers. The solid green and dashed pink curves refer to the results obtained by, respectively, Eqs.~(\ref{Eq41}) and (\ref{Eq33}). Here the system parameter is chosen as $G_0=0.2\omega_m$ and other parameters are the same as in Fig.~\ref{Fig2}.}\label{Fig7}
\end{figure}

In Fig.~\ref{Fig7}, we compare the steady-state position variance $\langle\delta\widetilde{X}_b^2\rangle_{\mathrm{s}}$ of the mechanical mode obtained from, respectively, the exact numerical solution in Eq.~(\ref{Eq41}) and the approximate analytical solution in Eq.~(\ref{Eq33}) with different mechanical bath mean phonon numbers. As confirmed in Fig.~\ref{Fig7}, the analytical solution under the adiabatic approximation agrees very well with the exact numerical result.

Once the analytical solution of the steady-sate position variance $\langle\delta\widetilde{X}_b^2\rangle_{\mathrm{s}}$ is obtained, the analytical optimal ratio of $G_1/G_0$ to maximize $\langle\delta\widetilde{X}_b^2\rangle_{\mathrm{s}}$ can be evaluated accordingly in principle by
\begin{eqnarray}\label{Eq42}
\frac{d\langle\delta\widetilde{X}_b^2\rangle_{\mathrm{s}}}{d(G_1/G_0)}=0.
\end{eqnarray}
After some simplifications, the optimal $G_1/G_0$ fulfills
\begin{eqnarray}\label{Eq43}
&&(1+2n_m)\frac{G_1}{G_0}\bigg|_{\mathrm{opt}}-\mathcal{C}\Big[1-\Big(\frac{G_1}{G_0}\bigg|_{\mathrm{opt}}\Big)^2\Big] \cr\cr
&&\times e^{-2\mathrm{arctanh}\frac{G_1}{G_0}\big|_{\mathrm{opt}}}=0,
\end{eqnarray}
which is a transcendental equation about $(G_1/G_0)|_{\mathrm{opt}}$ and whose analytical solution is hard to solve. However, if we further make approximation in the large enough cooperativity ($\mathcal{C}\gg1$)
\begin{eqnarray}\label{Eq44}
e^{-2r}\simeq\frac12\sqrt{\frac{1+2n_m}{\mathcal{C}}},
\end{eqnarray}
the optimal $G_1/G_0$ can be obtained analytically
\begin{eqnarray}\label{Eq45}
\frac{G_1}{G_0}\bigg|_{\mathrm{opt}}\simeq\sqrt{1+\frac{1+2n_m}{\mathcal{C}}}-\sqrt{\frac{1+2n_m}{\mathcal{C}}}.
\end{eqnarray}

\begin{figure}
\centering
\includegraphics[width=0.7\linewidth]{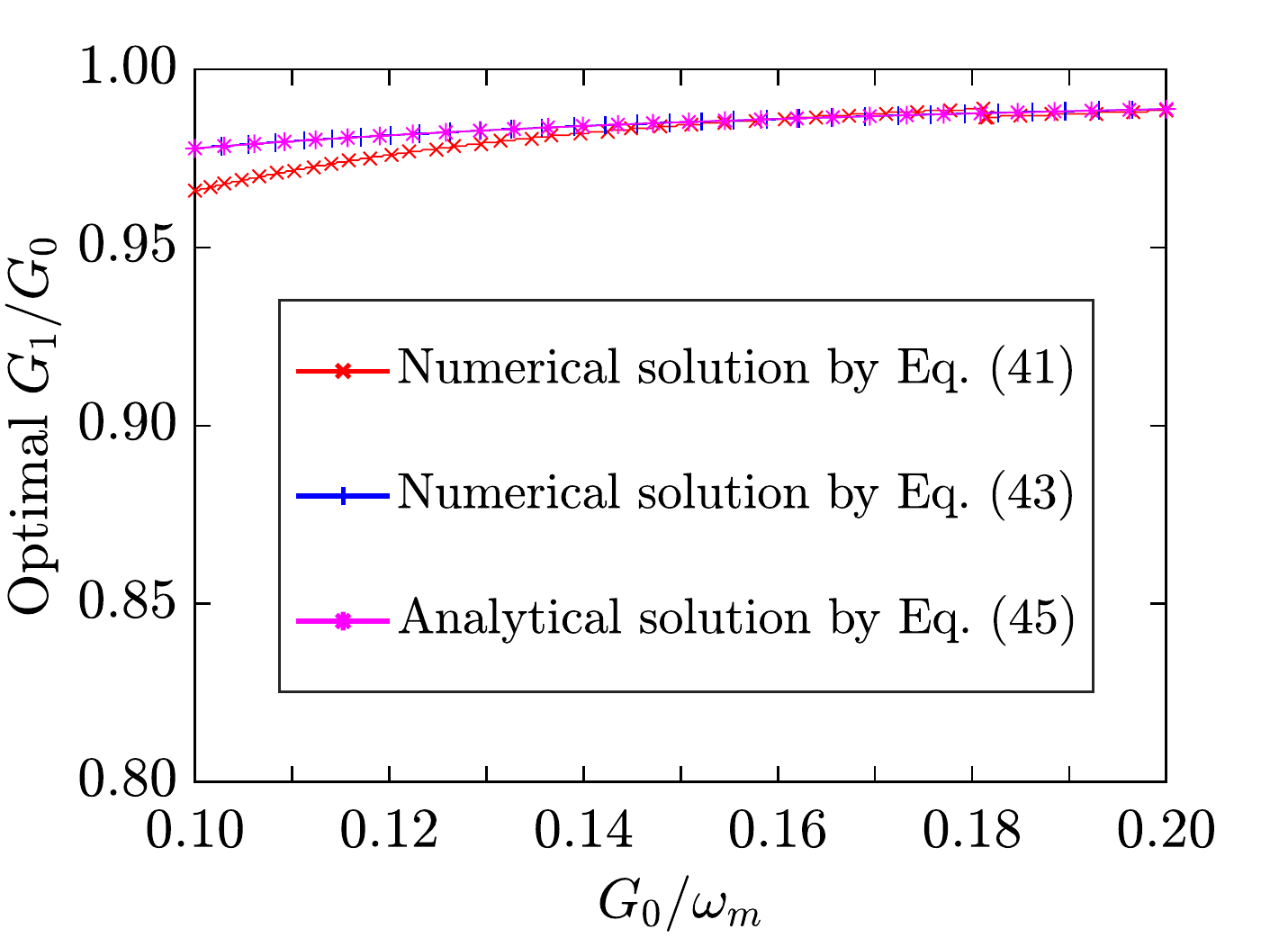}
\caption{(Color online) The optimal ratio $G_1/G_0$ versus the effective optomechanical coupling center sideband strength $G_0$, where $(G_1/G_0)|_{\mathrm{opt}}$ is evaluated with, respectively, numerical solution of Eq.~(\ref{Eq41}), numerical solution of Eq.~(\ref{Eq43}), and analytical solution of Eq.~(\ref{Eq45}). Here $n_m=100$ and other parameters are the same as in Fig.~\ref{Fig2}.}\label{Fig8}
\end{figure}

In Fig.~\ref{Fig8}, we plot the optimal $G_1/G_0$ as functions of $G_0$ with different methods, i.e., numerical solution of Eq.~(\ref{Eq41}), numerical solution of Eq.~(\ref{Eq43}), and analytical solution of Eq.~(\ref{Eq45}). One can note that there is only a little discrepancy among these results initially and they all converge together finally. Therefore, the analytical solution in Eq.~(\ref{Eq45}) is approximately valid.

\begin{figure}
\centering
\includegraphics[width=0.7\linewidth]{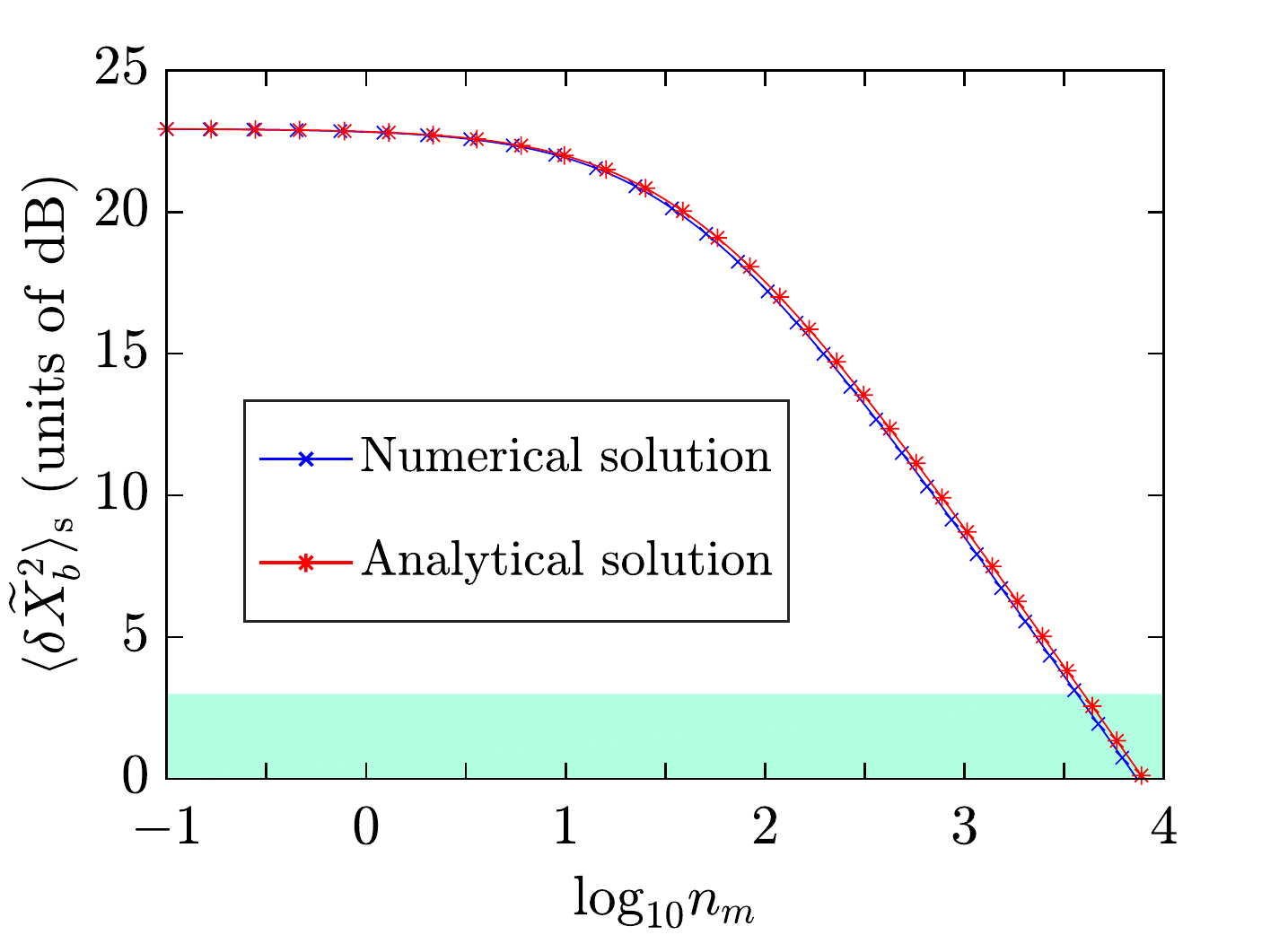}
\caption{(Color online) The position variance $\langle\delta\widetilde{X}_b^2\rangle_{\mathrm{s}}$ versus the thermal phonon occupation number $n_m$. The shadowed green bottom region corresponds to mechanical squeezing below the 3 dB limit. Here $\gamma_m=0.5\times10^{-6}\omega_m$, $G_0=0.1\omega_m$, and $G_1=0.99G_0$. Other system parameters are the same as in Fig.~\ref{Fig2}.}\label{Fig9}
\end{figure}

To further show the robustness of the mechanical squeezing engineered via the present method against the mechanical thermal noise, we plot the steady-state position variance $\langle\delta\widetilde{X}_b^2\rangle_{\mathrm{s}}$ obtained by, respectively, the numerical solution in Eq.~(\ref{Eq41}) and analytical solution in Eq.~(\ref{Eq33}), as a function of the thermal phonon occupation number $n_m$. As demonstrated in Fig.~\ref{Fig9}, when the bath temperature is low ($n_m\sim10$), the far beyond 3-dB limit strong mechanical squeezing ($\sim$ 22 dB) is achievable. The result also shows that the engineered squeezing has strong robustness. Even at a high bath temperature with $n_m\sim3\times10^3$, the steady-state mechanical squeezing can still break the 3-dB limit. Additionally, one can clearly note that the analytical result is in excellent agreement with the numerical calculation.

Before concluding, we briefly discuss the experimental feasibility about our mechanical squeezing scheme. In present scheme, the optomechanical setup used is a standard optomechanical cavity and it is significantly common in current cavity optomechanics~\cite{2014RMP861391}. The required system parameters are also in accessible range for existing optomechanical experiments. The applied technique of periodically modulating driving field has been highly mature until now, which is widely used to manipulate the optomechanical (electromechanical) systems~\cite{2009PRL103213603,2012PRA86013820,2012NJP14075014,2012NJP14125005}. Therefore, our squeezing scheme is remarkably workable with the nowaday optomechanics techniques.

\section{Conclusions}\label{Sec6}
In conclusion, we have proposed a simple but very effective method to engineer far surpassing 3-dB strong mechanical squeezing in a standard optomechanical system which only contains a cavity mode and a mechanical mode. The introduction of the suitable periodic modulation into the amplitude of the single-tone driving field enables us to obtain the desired form of the effective optomechanical coupling, which just contributes to cooling the Bogoliubov mode of the mechanical mode close to its ground state resorting to the interaction with the cavity mode. We analyze the role of the nonresonant terms produced by the periodically modulated effective optomechanical coupling playing in engineering squeezing and find that it leads to the continuous $\tau$-periodicity rotation of the direction of quadrature squeezing in the phase space. We demonstrate that the squeezing degree is not simply rely on the magnitude of the effective optomechanical coupling but closely on the sideband strength ratio $G_1/G_0$. It is shown that the engineered squeezing is a nonmonotonic function of $G_1/G_0$. The maximized squeezing is the result that the optimized $G_1/G_0$ arranges the competing effect between the squeezing of the mechanical mode and the cooling of the Bogoliubov mode to best tradeoff. In the steady-state regime, we both maximize the squeezing and optimize the ratio $G_1/G_0$ numerically and analytically, which are agree very well each other. We also show that the engineered squeezing has strong robustness against thermal noise and the periodic effective optomechanical coupling form required in our scheme can be precisely prepared via the explicit external single-tone driving field, which indicate that the present scheme is significantly feasible with available experimental platform in current cavity optomechanics. Compared with previous schemes, our scheme not only involves fewer control laser source, but also can be expected to simplify some existing schemes based on the two-tone pump driving technique.

\section*{ACKNOWLEDGMENTS}
This work was supported by the National Natural Science Foundation of China under Grant
(61822114, 61575055, 11465020, 61465013); The Project of Jilin Science and Technology Development for Leading Talent of Science and Technology Innovation in Middle and Young and Team Project under Grant (20160519022JH).

\appendix
\section{Asymptotic evolution of the amplitudes of the cavity and mechanical modes}\label{App1}
In the main text, we illustrated that when the performed external periodic driving is set as $\varepsilon_L(t)=\varepsilon_{-1}e^{i\Omega t}+\varepsilon_0+\varepsilon_1e^{-i\Omega t}$, the amplitudes of the cavity and mechanical modes will evolve toward a same structure in the long-time limit. To gain more insights about this kind of asymptotic process in dynamics clearly, we verify it here from the point of the analytical expressions.

In the parameter regime of $g_0\ll\omega_m$, the optomechanical coupling coefficient $g_0$ in Eq.~(\ref{Eq4}) can be treated as the perturbation. Meanwhile, due to the periodicity of the implemented driving [$\varepsilon_L(t)=\varepsilon_L(t+\tau)$], the asymptotic amplitudes $\langle a(t)\rangle$ and $\langle b(t)\rangle$ will be also $\tau$-periodic. Therefore, the asymptotic solutions of Eq.~(\ref{Eq4}) can be made the double expansions (perturbation expansion and Fourier expansion):
\begin{eqnarray}\label{EqA1}
\langle\mathscr{O}(t)\rangle=\sum_{j=0}^{\infty}\sum_{n=-\infty}^{\infty}\mathscr{O}_{n,j}e^{in\Omega t}g_0^j~~(\mathscr{O}=a, b),
\end{eqnarray}
where the expansion coefficient $\mathscr{O}_{n,j}$ is time-independent. Substituting above equation into Eq.~(\ref{Eq4}), the zeroth-order perturbation coefficients can be obtained
\begin{eqnarray}\label{EqA2}
a_{n,0}=\frac{E_{-n}}{i(\delta_a+n\Omega)+\frac{\kappa}{2}},~~~b_{n,0}=0,
\end{eqnarray}
and the $j$th-order perturbation coefficients ($j\geqslant1$) can be also gained in the following way of recursive relations
\begin{eqnarray}\label{EqA3}
a_{n,j}&=&i\sum_{k=0}^{j-1}\sum_{m=-\infty}^{\infty}\frac{a_{n+m,j-k-1}b_{m,k}^{\ast}+a_{n-m,j-k-1}b_{m,k}}{i(\delta_a+n\Omega)+\frac{\kappa}{2}}, \cr\cr
b_{n,j}&=&i\sum_{k=0}^{j-1}\sum_{m=-\infty}^{\infty}\frac{a_{n+m,j-k-1}a_{m,k}^{\ast}}{i(\omega_m+n\Omega)+\frac{\gamma_m}{2}}.
\end{eqnarray}
Therefore, the sideband amplitudes $\mathscr{O}_n$ in Eq.~(\ref{Eq6}) can be expressed as
\begin{eqnarray}\label{EqA4}
\mathscr{O}_n=\sum_{j=0}^{\infty}\mathscr{O}_{-n,j}g_0^j~~(n=-1, 0, 1).
\end{eqnarray}
\begin{figure}
\centering
\includegraphics[width=1.0\linewidth]{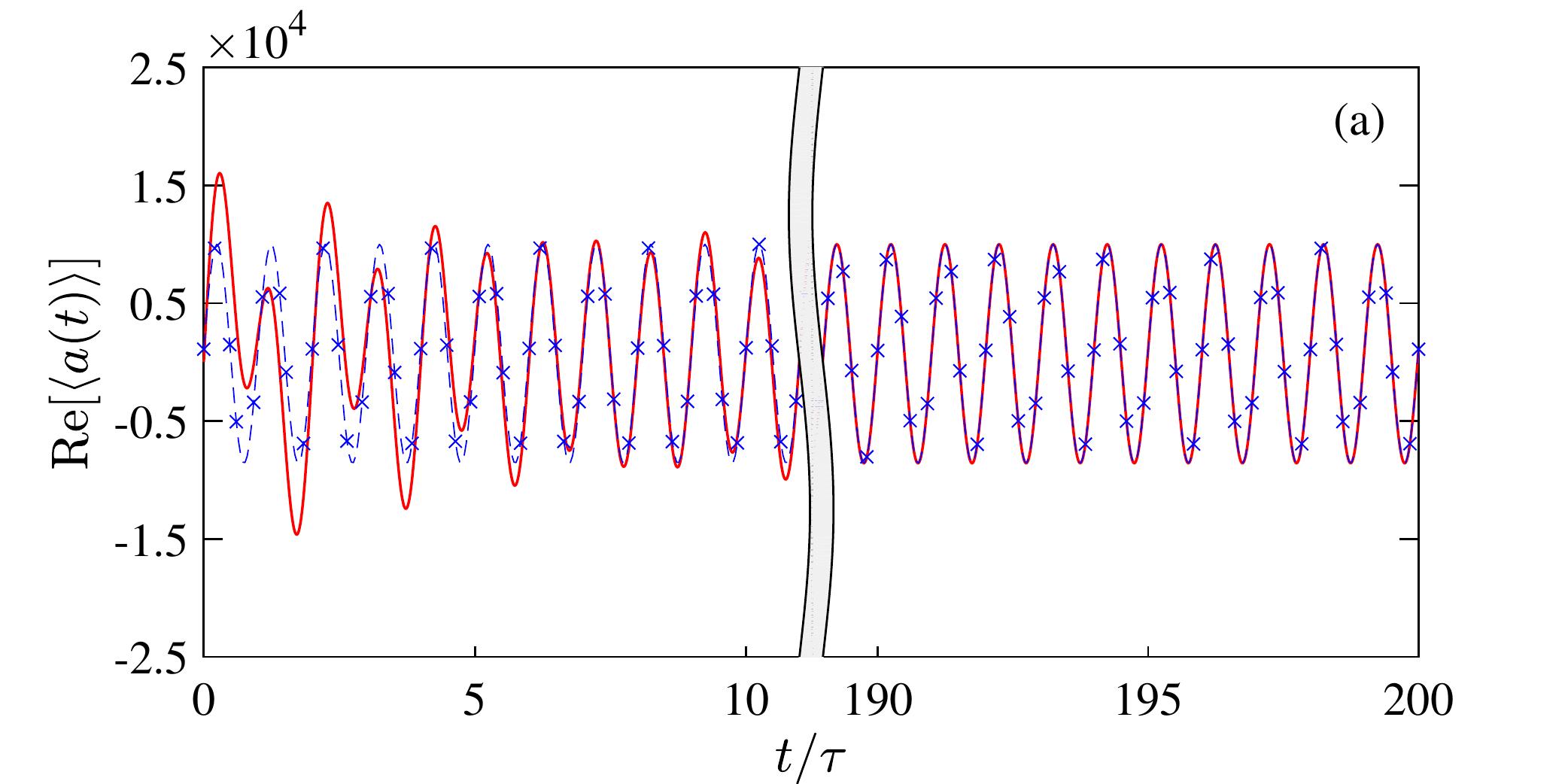}
\includegraphics[width=1.0\linewidth]{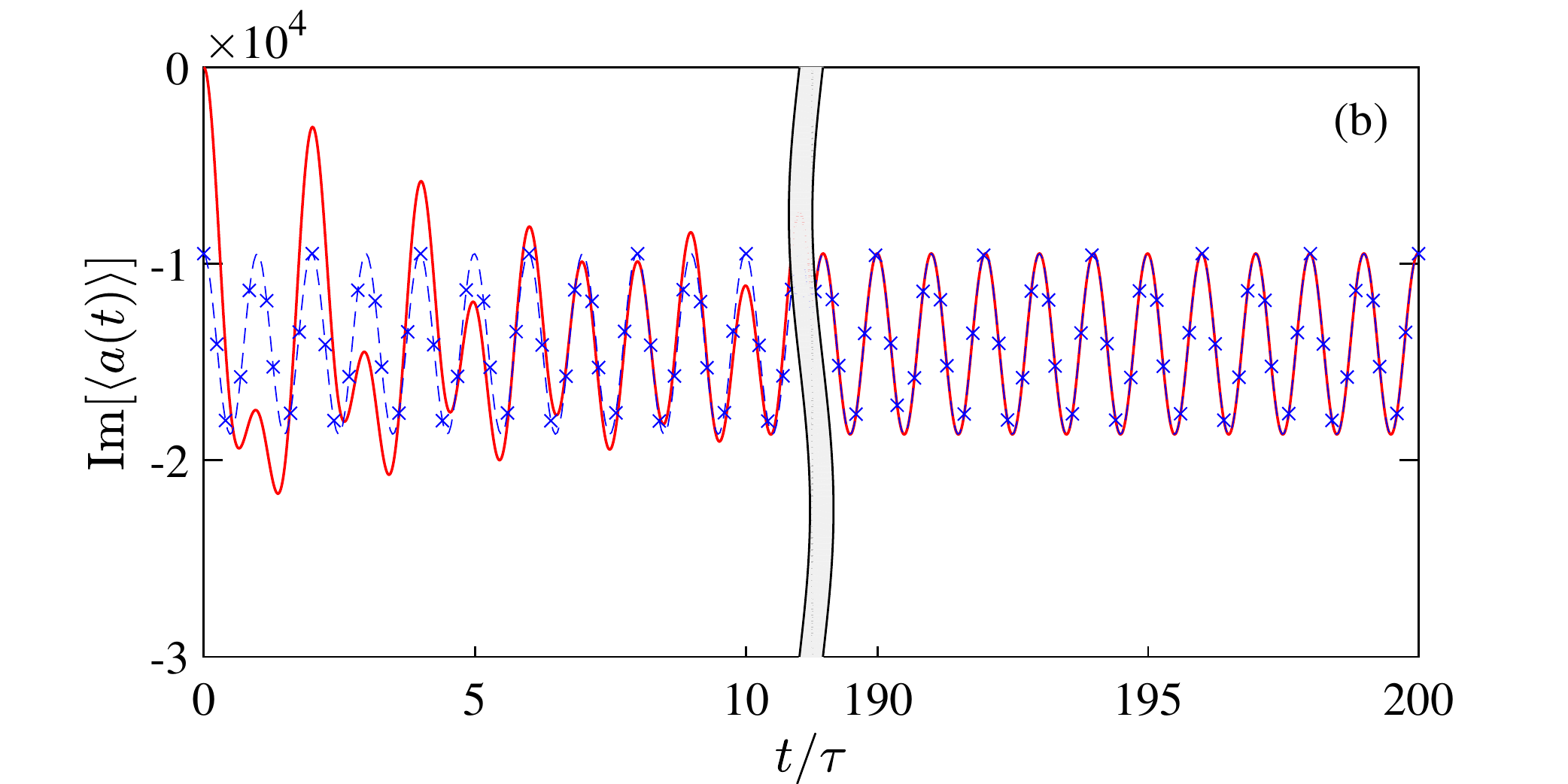}
\includegraphics[width=1.0\linewidth]{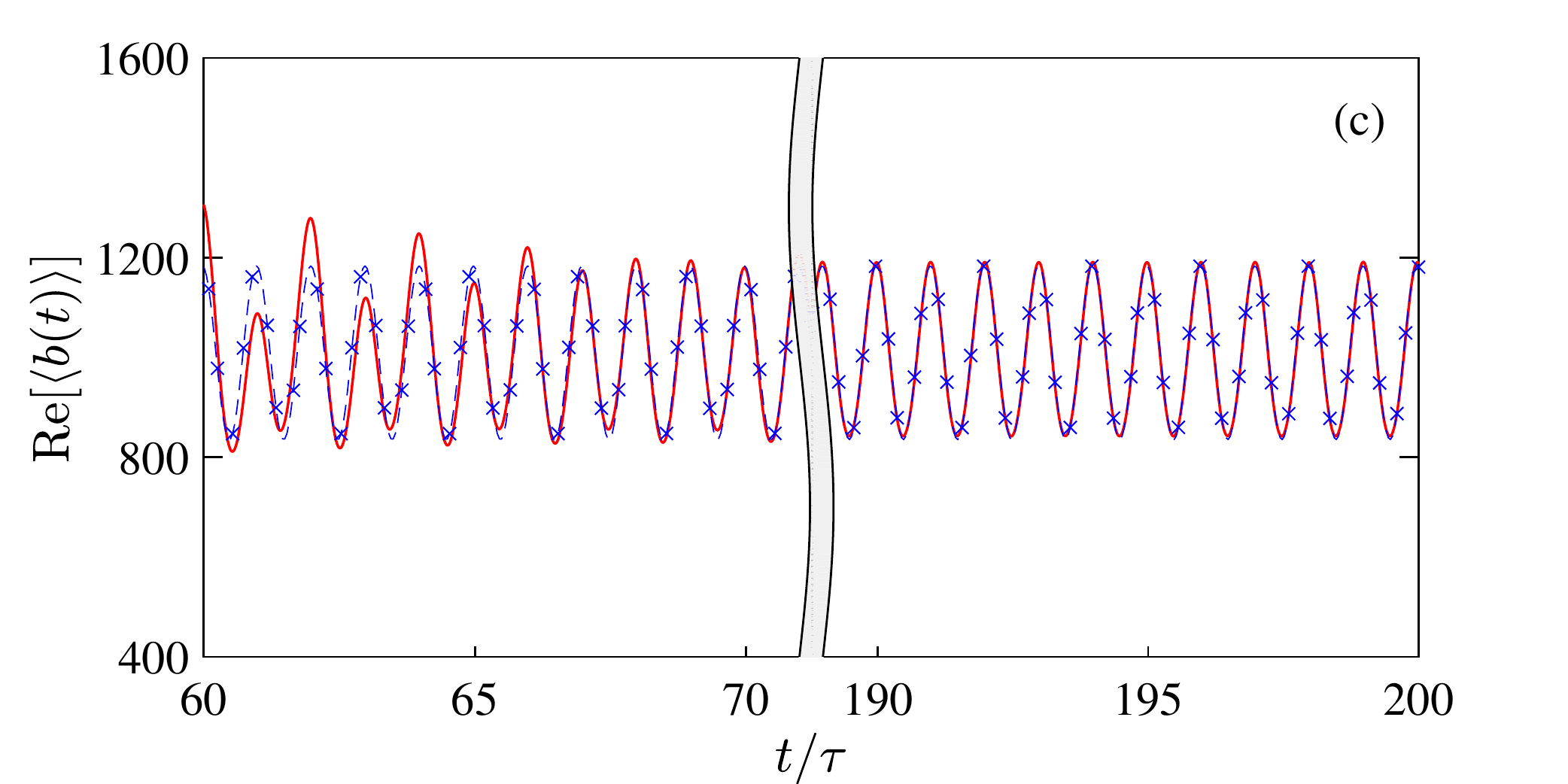}
\includegraphics[width=1.0\linewidth]{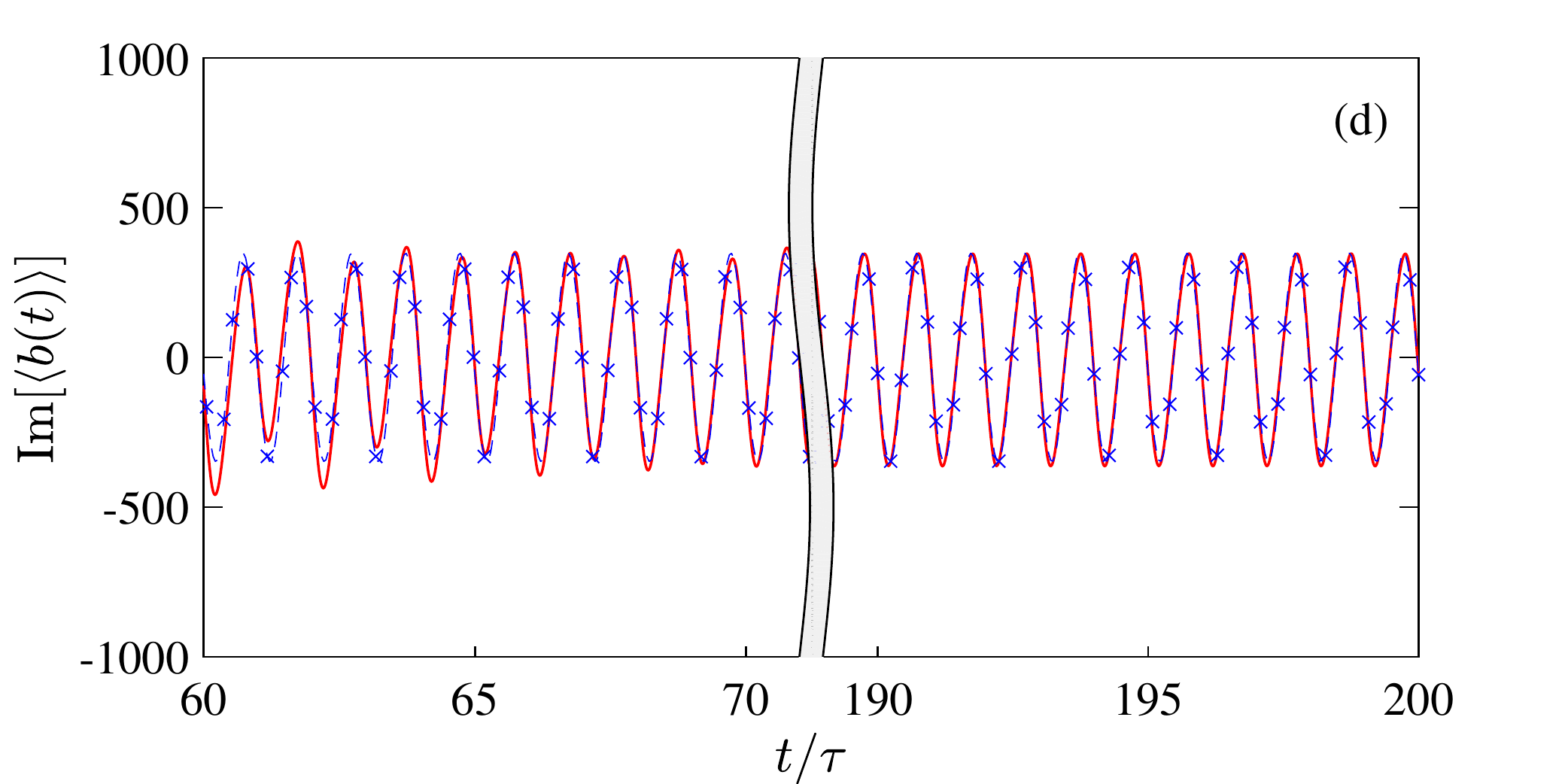}
\caption{(Color online) Asymptotic time evolution of the system amplitudes in the modulation periods of $[0, 200\tau]$. Real and imaginary parts of the cavity mode amplitude $\langle a(t)\rangle$ (mechanical mode amplitude $\langle b(t)\rangle$) versus modulation time $t$, respectively, in (a) and (b) ((c) and (d)). In all figures, the red solid and blue dashed cross lines are the results obtained from, respectively, the numerical solution in Eq.~(\ref{Eq4}) and analytical expression in Eq.~(\ref{Eq6}). The system parameters are chosen as (in units of $\omega_m$): $\gamma_m=10^{-6}$, $\delta_a=1$, $\kappa=0.1$, $g_0=4\times10^{-6}$, $\varepsilon_0=1.4\times10^4$, and $\varepsilon_{\pm1}=0.7\times10^4$.}\label{FigA1}
\end{figure}

To verify the validity of the structure as shown in Eq.~(\ref{Eq6}) in the long-time limit, in Fig.~\ref{FigA1}, we explicitly present the time evolution of the system amplitudes $\langle a(t)\rangle$ and $\langle b(t)\rangle$ in the modulation periods of $[0, 200\tau]$ with the exact numerical solution of Eq.~(\ref{Eq4}) and the analytical expression of Eq.~(\ref{Eq6}), respectively. The first half of each subfigure in Fig.~\ref{FigA1} clearly exhibits the slow approaching process in dynamics between these two different kinds of results. While from the second part of them, it is fantastically found that these two kinds of results converge together perfectly in the long-time modulation periods ($[190\tau, 200\tau]$). Therefore, the cavity mode amplitude $\langle a(t)\rangle$ and the mechanical mode amplitude $\langle b(t)\rangle$ do indeed have the same structure with the external performed driving modulation and are $\tau$ periodic in the long-time limit. Moreover, from Fig.~\ref{FigA1}, one can also clearly find that the $\tau$-periodic asymptotic process of the cavity mode amplitude is much faster than that of the mechanical mode amplitude. This is because the external periodic driving is directly performed on the cavity mode while the asymptotic $\tau$ periodicity of the mechanical mode is obtained via the intermediate mode (cavity mode) based on the optomechanical interaction.

Here we should point out that, for gaining high enough level of approximation, it has been truncated the perturbation series in Eq.~(\ref{EqA4}) up to $j\leq10$ during calculating the analytical solution.

\section{Choice of the sideband-modulation strengths for external driving $\varepsilon_L(t)$ to fulfill the desired $G(t)$}\label{App2}
\begin{figure}
\centering
\includegraphics[width=1.0\linewidth]{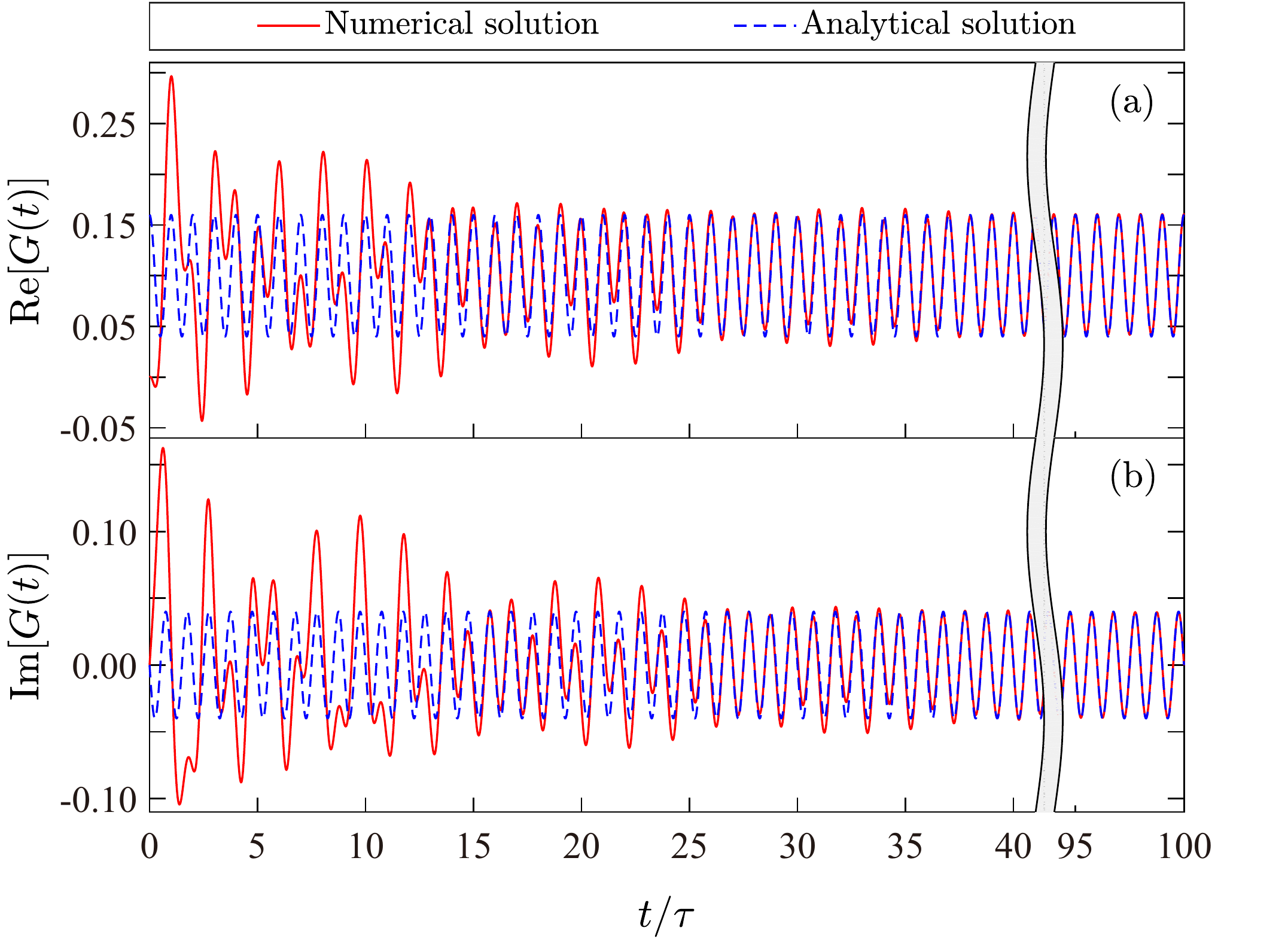}
\caption{(Color online) Asymptotic time evolution of real and imaginary parts (in units of $\omega_m$) for the effective optomechanical coupling $G(t)$ in (a) and (b), respectively. The solid red curve shows the numerical result when the external periodic driving $\varepsilon_L(t)=\varepsilon_{-1}e^{i\Omega t}+\varepsilon_0+\varepsilon_1e^{-i\Omega t}$ is acted on the present optomechanical system, where the corresponding set of sideband-modulation strengths $(\varepsilon_{-1}, \varepsilon_0, \varepsilon_1)$ is given in Eq.~(\ref{EqA5}). While the dashed blue curve represents the analytical result obtained by the assumed $G(t)$ in Eq.~(\ref{Eq7}). The system parameters are as follows (in units of $\omega_m$): $\gamma_m=10^{-6}$, $\delta_a=1$, $\kappa=0.1$, $g_0=4\times10^{-6}$, $G_{-1}=0.01$, $G_0=0.1$, and $G_1=0.05$.}\label{FigA2}
\end{figure}

To fulfill the desired form of the effective optomechanical coupling $G(t)$ in Eq.~(\ref{Eq7}) when a set of $(G_{-1}, G_0, G_1)$ is given, the corresponding sideband-modulation strengths $(\varepsilon_{-1}, \varepsilon_0, \varepsilon_1)$ for the external driving $\varepsilon_L(t)$ can be derived analytically via Laplace transform:
\begin{eqnarray}\label{EqA5}
\varepsilon_{-1}&=&\frac{G_{-1}}{g_0}\left[i(\Omega+\delta_a)+\frac{\kappa}{2}\right] \cr\cr
&&-i\left[2k_0G_{-1}+(k_3+k_4)G_0+(k_1+k_2)G_1\right], \cr\cr
\varepsilon_0&=&\frac{G_0}{g_0}(i\delta_a+\frac{\kappa}{2}) \cr\cr
&&-i\left[(k_3+k_4)G_{-1}+2k_0G_0+(k_3+k_4)G_1\right], \cr\cr
\varepsilon_1&=&\frac{G_1}{g_0}\left[i(\delta_a-\Omega)+\frac{\kappa}{2}\right] \cr\cr
&&-i\left[(k_1+k_2)G_{-1}+(k_3+k_4)G_0+2k_0G_1\right], \cr\cr
&&
\end{eqnarray}
where
\begin{eqnarray}\label{EqA6}
k_0&=&-\frac{i(G_{-1}^2+G_0^2+G_1^2)}{2g_0S_1}, \cr\cr
k_1&=&-\frac{iG_{-1}G_1S_2}{g_0(S_1-S_2)(S_2-S_3)}, \cr\cr
k_2&=&\frac{iG_{-1}G_1S_3}{g_0(S_1-S_3)(S_2-S_3)}, \cr\cr
k_3&=&-\frac{iG_0(G_{-1}+G_1)S_4}{g_0(S_1-S_4)(S_4-S_5)}, \cr\cr
k_4&=&\frac{iG_0(G_{-1}+G_1)S_5}{g_0(S_1-S_5)(S_4-S_5)}, \cr\cr
S_1&=&-i\omega_m-\frac{\gamma_m}{2}, \cr\cr
S_2&=&2i\Omega,~~~S_3=-2i\Omega, \cr\cr
S_4&=&i\Omega,~~~~S_5=-i\Omega.
\end{eqnarray}

To check the validity of the above derived external periodic driving $\varepsilon_L(t)$, in Fig.~\ref{FigA2}, we compare the effective optomechanical coupling $G(t)$ obtained by, respectively, the numerical solution when $\varepsilon_L(t)=\varepsilon_{-1}e^{i\Omega t}+\varepsilon_0+\varepsilon_1e^{-i\Omega t}$ is applied to the present optomechanical system and the analytical solution assumed in Eq.~(\ref{Eq7}). From Fig.~\ref{FigA2}, one can note that these two different kinds of solutions agree very well in the modulation periods of $[95\tau, 100\tau]$. It means that the assumed effective optomechanical coupling form in Eq.~(\ref{Eq7}) which is desired and necessary in generation of mechanical squeezing can be precisely engineered in the long-time modulation limit via choosing the suitable external driving-sideband strengths given in Eq.~(\ref{EqA5}).



%

\end{document}